\documentclass[11pt,a4paper]{article}
\usepackage{epsfig,amsmath,amssymb,scalefnt,cite,mcite}
\usepackage{xcolor} %take out at the end

\long\def\symbolfootnote[#1]#2{\begingroup%
\def\thefootnote{\fnsymbol{footnote}}\footnote[#1]{#2}\endgroup}

\let\OLDthebibliography\thebibliography
\renewcommand\thebibliography[1]{
  \OLDthebibliography{#1}
  \setlength{\parskip}{0pt}
  \setlength{\itemsep}{0pt plus 0.3ex}
}

\newcommand{\abbrev}{\rm\scalefont{.9}}

\newcommand{\sm}{{\abbrev SM}}
\newcommand{\thdm}{{\abbrev 2HDM}}

\newcommand{\cp}{$\mathcal{CP}$}

\newcommand{\nlo}{{\abbrev NLO}}
\newcommand{\lo}{{\abbrev LO}}
\newcommand{\qcd}{{\abbrev QCD}}
\newcommand{\lhc}{{\abbrev LHC}}

\newcommand{\smallz}{{\scriptscriptstyle Z}} %  a smaller Z
\newcommand{\smallw}{{\scriptscriptstyle W}} %
\newcommand{\smallH}{{\scriptscriptstyle H}} %
\newcommand{\mz}{m_\smallz}

\newcommand{\mzz}{m_{\smallz\smallz}}

\newcommand{\mw}{m_\smallw}
\newcommand{\mH}{m_\smallH}

\newcommand{\GaH}{\Gamma_\smallH}
\newcommand{\Gaz}{\Gamma_\smallz}
\newcommand{\Gaw}{\Gamma_\smallw}

\newcommand{\cba}{c_{\beta-\alpha}}
\newcommand{\sba}{s_{\beta-\alpha}}
\newcommand{\tb}{t_{\beta}}

\newcommand{\eqn}[1]{Eq.\,(\ref{#1})}

\newcommand{\fig}[1]{Fig.\,\ref{#1}}

\newcommand{\tab}[1]{Tab.\,\ref{#1}}
\newcommand{\sct}[1]{Section~\ref{#1}}

\newcommand{\citere}[1]{Ref.~\cite{#1}}
\newcommand{\citeres}[1]{Refs.~\cite{#1}}

\begin{document}

\begin{titlepage}

{\flushright{
        \begin{minipage}{5cm}
          DESY 15-239, ZU-TH 43/15
        \end{minipage}        }

}
\renewcommand{\thefootnote}{\fnsymbol{footnote}}
\vskip 2cm
\begin{center}
{\LARGE\bf 
Interference contributions to gluon initiated heavy\\[2mm] Higgs production in the Two-Higgs-Doublet Model
}
\vskip 1.0cm
{\Large  Nicolas Greiner$^{a,b}$, Stefan Liebler$^{a}$, Georg Weiglein$^{a}$}
\vspace*{8mm} \\
{\sl ${}^a$
DESY, Notkestra\ss e 85, \\
22607 Hamburg, Germany}
\vspace*{8mm} \\
{\sl ${}^b$
Physik-Institut, Universit\"at Z\"urich, Winterthurerstrasse 190\\
8057 Z\"urich, Switzerland}
\end{center}
\symbolfootnote[0]{{\tt emails: greiner@physik.uzh.ch, stefan.liebler@desy.de, georg.weiglein@desy.de}}

\vskip 0.7cm

\begin{abstract}
We discuss the production of a heavy neutral Higgs boson 
of a \cp{}-conserving Two-Higgs-Doublet Model in gluon fusion and its decay
into a four-fermion final state,
$gg (\rightarrow VV) \rightarrow e^+e^-\mu^+\mu^-/e^+e^-\nu_l\bar\nu_l$.
We investigate the interference contributions to invariant mass
distributions of the four-fermion final state and other relevant kinematical
observables. The relative importance of the different contributions is
quantified for the process in the on-shell approximation,
$gg\rightarrow ZZ$.
We show that interferences
of the heavy Higgs with the light Higgs boson and background contributions 
are essential
for a correct description of the differential cross section. 
Even though they contribute below $\mathcal{O}(10\%)$ to those
heavy Higgs signal
cross sections, to which the experiments at the Large Hadron Collider
were sensitive in its first run, we find that
they are sizeable in certain regions of the parameter space that are
relevant for future heavy Higgs boson searches. 
In fact, the interference contributions 
can significantly enhance the experimental sensitivity
to the heavy Higgs boson.
\end{abstract}
\vfill
\end{titlepage}    

\setcounter{footnote}{0}

%%%%%%%%%%%%%%%%%%%%%%%%%%%%%%%%%%%%%%%%%%%%%%%%%%%%%%%%%%%%%%%%%%%%%%%%%%%%

\section{Introduction}
\label{sec:intro}

The two multi-purpose experiments ATLAS and CMS at the CERN Large Hadron
Collider (\lhc{}) discovered in 2012 a scalar resonance at
$125$\,GeV~\cite{Aad:2012tfa,Chatrchyan:2012ufa}, which
is compatible with a Standard Model (\sm{}) Higgs boson.
Even though its couplings are --- to the precision obtained so far ---
in agreement with the \sm{} expectations, it can well be embedded in
an extended Higgs sector like a Two-Higgs-Doublet Model (\thdm{}).

The main production mechanism for a \sm{}-like Higgs boson~$h$ is gluon fusion~\cite{Georgi:1977gs},
for which a large amount of higher-order corrections
in quantum chromodynamics (\qcd{})~\cite{Dawson:1990zj,Djouadi:1991tka,
Graudenz:1992pv,Spira:1995rr,Harlander:2002wh,Anastasiou:2002yz,Ravindran:2003um,Anastasiou:2015ema}
are known.
Recently the combined scale and PDF$+\alpha_s$ uncertainties, see e.g.\
\citeres{Dittmaier:2011ti,Dittmaier:2012vm,Heinemeyer:2013tqa}, were
reduced to below $\mathcal{O}(10)$\%.
Of particular importance for the discovery of the new particle and the
subsequent investigations
of its mass and couplings were
the decays into heavy gauge bosons $h\rightarrow VV$
with $V\in \lbrace{W,Z\rbrace}$, which appear to be of relevance also
for off-shell Higgs bosons~\cite{Kauer:2012hd} (see also 
\citere{Liebler:2015aka} for an investigation of off-shell contributions at
a linear collider and the \lhc).
Given the small theory uncertainties and because of their importance
for the unitarization of the process, interference effects between the 
off-shell Higgs boson
and the continuum background in $gg\rightarrow h^*\rightarrow VV$ were studied for
leptonic decays in \citeres{Glover:1988fe,Glover:1988rg,Binoth:2006mf,
Campbell:2011cu,Kauer:2012ma,Passarino:2012ri,Cascioli:2013gfa,
Kauer:2013qba,Campbell:2013una,Ellis:2014yca,Campanario:2012bh,Campbell:2014gua,Bonvini:2013jha,Li:2015jva}.
The semileptonic process, where interferences with tree-level background diagrams occur,
was recently discussed for the first time \cite{Kauer:2015dma}.
Whereas for the processes $gg\rightarrow hh$, 
$gg\rightarrow hZ$ and $gg\rightarrow\gamma\gamma$
next-to-leading order (\nlo{}) \qcd{} contributions have been known for quite some time~\cite{Dawson:1998py,Altenkamp:2012sx,Bern:2002jx},
the calculation of background contributions to $gg\rightarrow VV$ for off-shell gauge bosons in the limit of
massless quarks at \nlo{} \qcd{} became available just very recently~\cite{Caola:2015ila,vonManteuffel:2015msa,Melnikov:2015laa,Caola:2015rqy}.
For the mentioned signal-background interference approximate higher order contributions using a soft-collinear approximation
for a heavy Higgs boson~\cite{Bonvini:2013jha} 
and applying soft-gluon resummation for an off-shell light Higgs boson~\cite{Li:2015jva} were previously available.
 
We discuss 
$gg\rightarrow (VV) \rightarrow e^+e^-\mu^+\mu^-/e^+e^-\nu_l\bar\nu_l$
at \lo{} \qcd{} in the context of a \cp{}-conserving \thdm{}, see
\citeres{Gunion:1989we,Akeroyd:1996he,Akeroyd:1998ui,Aoki:2009ha,Branco:2011iw,Craig:2012vn} for \thdm{} reviews.
We furthermore assume 
the absence of tree-level flavor-changing neutral currents.
The Higgs sector of the \cp-conserving \thdm{} consists of three Higgs 
bosons $\phi_i,i\in\lbrace 1,2,3\rbrace$, namely
two \cp{}-even Higgs bosons $h,H$ 
with masses $m_h<\mH$ and one \cp{}-odd Higgs boson $A$.
According to the structure of the Yukawa couplings four types of \thdm{}s are
distinguished, where for our purposes only two types are of relevance,
namely those with different couplings of the two Higgs doublets to up- and down-type quarks.
The process $gg\rightarrow VV$ was only 
very recently discussed in the context of a \thdm{}~\cite{Jung:2015sna},
following a discussion of $gg\rightarrow t\bar t$~\cite{Jung:2015gta}.
In contrast, for the similar processes
$gg\rightarrow \phi_i \phi_j$~\cite{Hespel:2014sla}
and $gg\rightarrow Z\phi_i$~\cite{Hespel:2015zea} in the \thdm{} even higher order
effects were partially included.
The process $gg\rightarrow VV$ was recently extensively discussed
in the context of the \sm{} with
an additional real singlet~\cite{Maina:2015ela,Kauer:2015hia}, and also
the vector-boson fusion process was considered
in \citere{Ballestrero:2015jca}.
The extension of the \sm\ by a real singlet can be characterised by a
single angle which multiplies
all Higgs couplings to quarks, leptons and gauge bosons with a
universal factor.
Accordingly the phenomenology of the \thdm{} is more rich, since 
in the \thdm\ in particular the couplings
of the two \cp{}-even Higgs bosons to quarks and to gauge bosons are
modified differently.

The interference contributions to the process 
$gg\rightarrow (VV) \rightarrow e^+e^-\mu^+\mu^-/e^+e^-\nu_l\bar\nu_l$,
within the \thdm, which we study in the present paper, are of interest for
several reasons. The interference
contributions of the heavy Higgs with the light Higgs and the background are
crucial for the
unitarization of the process. The related effects are particularly important 
for high invariant masses of the gauge bosons, i.e.\ at high energies of the
hard scattering process. In general, the interference effects need to be 
well understood in order to obtain a sufficiently accurate prediction for the 
process. Furthermore, interference effects are of interest since they 
can potentially enhance the sensitivity to the signal of a heavy Higgs
boson. All those aspects are addressed in our analysis below.

We make use of {\tt GoSam}~\cite{Cullen:2011ac,Cullen:2014yla}
to discuss the processes $gg\rightarrow e^+e^-\mu^+\mu^-$ and
$e^+e^-\nu_l\bar\nu_{l}$ (including all three neutrino flavors).
For a study of the relevance of interference contributions
we also consider the case where the first process is approximated by the
on-shell production of two $Z$~boson, $gg\rightarrow ZZ$ (and the subsequent
decays of the $Z$~bosons). We added its amplitudes for this process
to a modified version~\cite{Harlander:2013mla}
of {\tt vh@nnlo}~\cite{Brein:2012ne},
which has been linked to {\tt 2HDMC}~\cite{Eriksson:2009ws}.

The LHC experiments recently presented results for a heavy Higgs search with
subsequent decay into
heavy gauge bosons in \citeres{Khachatryan:2015cwa,Aad:2015kna}.
The ATLAS experiment also provided an interpretation in terms of the \thdm{}, neglecting
possible interferences between the heavy Higgs signal and the background
as well as with the contribution of
the light Higgs boson. The CMS experiment took into account a rescaled interference
from the \sm\ case of a heavy Higgs boson.
Neglecting the interference contributions involving the heavy Higgs boson
and employing the narrow-width approximation (NWA)
for the heavy Higgs boson, $gg\rightarrow H\rightarrow VV$, has of course 
the advantage that
all known QCD and electroweak corrections for gluon fusion and
the decay into heavy quarks as implemented in codes like {\tt SusHi}~\cite{Harlander:2012pb}
or {\tt Prophecy4F}~\cite{Bredenstein:2006ha,Bredenstein:2006rh} 
can be taken into account.
In our analysis of the interference contributions to the process
$gg\rightarrow (VV) \rightarrow e^+e^-\mu^+\mu^-/e^+e^-\nu_l\bar\nu_l$
we find that neglecting the interference contributions of the 
heavy Higgs boson with the background and the light Higgs boson in the ATLAS
analysis has indeed been justified in view of the experimental sensitivity
that has been reached in the first run of the \lhc. On the other hand, we
find that these interference contributions will be of relevance
for high integrated luminosities at the \lhc{}.

With respect to the interference contributions involving the heavy Higgs 
boson it is obviously important to ensure that the cross section into heavy
gauge bosons is correctly unitarized,
in particular at high invariant masses of the gauge boson system.
In case interferences involving the heavy Higgs boson are neglected
the light Higgs boson and its interference with the background has to
have \sm{}-like Higgs boson couplings. For the pure signal strength
of a heavy Higgs boson \citere{Jung:2015sna} suggests to use
a multiplicative factor covering the interference effects in the context 
of a \thdm{}. We find it preferable to take into account all interferences
of the light Higgs boson, the heavy Higgs boson and the background consistently in the
setup of a \thdm{} in order to describe the cross section at high invariant
masses accurately, as done in the present paper.
We find that 
in the vicinity of the heavy Higgs boson mass peak the role of interferences
is essential since they simultaneously alter the form and the position of the
heavy Higgs boson mass peak. We furthermore show that in certain regions of
the parameter space 
the interferences of the heavy Higgs boson with the light Higgs boson and the background
significantly enhance the heavy Higgs boson signal and thus increase the experimental sensitivity
in heavy Higgs boson searches.

The paper is organized as follows: We explain the
theoretical background in \sct{sec:theoback} starting
with a short introduction to the Two-Higgs-Doublet Model and providing
the details of our cross section calculations
for the processes under consideration.
In \sct{sec:paramchoice} we discuss the \thdm{} scenarios that
we consider in our study, and briefly describe the employed selection
cuts. Lastly, we present our numerical results in \sct{sec:results}
and conclude in \sct{sec:concl}.

\section{Theoretical background}
\label{sec:theoback}

In this section we first discuss 
the basics of the Two-Higgs-Doublet Model 
(\thdm{}) and afterwards
describe our implementation of
the processes $gg\rightarrow ZZ$ and $gg\rightarrow e^+e^-\mu^+\mu^-/e^+e^-\nu_l\bar\nu_l$
with the help of {\tt FeynArts}, {\tt FormCalc} and {\tt GoSam}.

\subsection{Two-Higgs-Doublet Model}
\label{sec:2HDM}
 
The \thdm{} contains two Higgs doublets, which we name $H_1$ and $H_2$.
It can conveniently be classified in four types if one demands the absence of
tree-level flavor-changing neutral currents
and furthermore assumes 
\cp{} conservation. By convention, the up-type quarks couple to $H_2$,
such that the couplings to the down-type quarks and to the leptons
can either be through $H_1$ or $H_2$, which corresponds to the four different types.
For details we refer to
\citeres{Gunion:1989we,Akeroyd:1996he,Akeroyd:1998ui,Aoki:2009ha,Branco:2011iw,Craig:2012vn}.
Since our studies are not sensitive to the coupling of the Higgs bosons
to leptons, it is sufficient to restrict our discussion to the
two types I and II.
The two Higgs doublets form one \cp{}-odd field $A$ and two
\cp{}-even Higgs fields $h$ and $H$ due to \cp{} conservation,
as well as two charged Higgs bosons $H^\pm$.
The \thdm{} can be described in different basis representations. We make use
of the ``physical basis'', in which the masses of all physical Higgs bosons,
the ratio of the vacuum expectation values $\tb:=\tan\beta=v_2/v_1$
and the Higgs mixing angle in the \cp{}-even sector $\alpha$, or alternatively
$\sba:=\sin(\beta-\alpha)$, are taken as input parameters.
Together with the mass term $m_{12}^2$ of both Higgs bosons $H_1^\dagger H_2$
all parameters of the Higgs sector of a \thdm{} are fixed.
We choose $\beta-\alpha$ in between $-\pi/2\leq \beta-\alpha\leq \pi/2$, such that $-1\leq \sba \leq 1$
and $0\leq \cba\leq 1$.
Our scenarios are thus specified by the two angles $\alpha$ and $\beta$,
which completely determine the relative couplings (with respect to the couplings
of a \sm{} Higgs boson) of the light and the heavy
Higgs boson to quarks and the heavy gauge boson. They are provided in \eqn{eq:gVV}
and \tab{tab:couplings} (together with \eqn{eq:gtbH} for a decomposition in
terms of $\beta-\alpha$ and $\beta$). Moreover, our analysis is sensitive to $m_h$ and $\mH$,
whereas it is rather insensitive to the mass of
the pseudoscalar $m_A$ and the heavy charged Higgs boson mass $m_{H^\pm}$,
as long as they are heavy enough not to open decay modes of the
heavy Higgs $H$ into them and as long as the decay mode $H\rightarrow hh$ is sub-dominant.
If the latter condition is fulfilled, also the dependence on the mass term $m_{12}^2$ can be neglected.

\begin{table}[htb]
\begin{center}
\begin{tabular}{| c || c | c |c| c|}
\hline
Model      & $g_u^h$ & $g_d^h$  & $g_u^H$ & $g_d^H$  \\\hline\hline
Type I     & $\cos\alpha/\sin\beta$ & $\cos\alpha/\sin\beta$ & $\sin\alpha/\sin\beta$ & $\sin\alpha/\sin\beta$\\\hline
Type II    & $\cos\alpha/\sin\beta$  & $-\sin\alpha/\cos\beta$  & $\sin\alpha/\sin\beta$  & $\cos\alpha/\cos\beta$ \\\hline
\end{tabular}
\end{center}
\vspace{-0.6cm}
\caption{Relative couplings $g_f^\phi$ (with respect to the \sm{} coupling) for the two \thdm{} types.}
\label{tab:couplings}
\end{table}

The Higgs boson couplings to the gauge bosons $V\in\lbrace W,Z\rbrace$ relative to the \sm{}
are given by
\begin{align}
 g_{V}^h = \sin(\beta-\alpha)=:\sba,\qquad g_{V}^{H}=\cos(\beta-\alpha)=:\cba\quad.
\label{eq:gVV}
\end{align}
The pseudoscalar has no lowest-order couplings to a pair of gauge bosons.
It can in principle contribute to the considered processes with four
fermions in the final state. Because of the suppression of the 
Yukawa couplings to leptons, however, these contributions are very small, 
and thus diagrams involving the pseudoscalar are
not of relevance for our discussion.
The relative couplings of the heavy
Higgs boson to bottom-quarks and top-quarks, which are 
of particular relevance for our discussion, are 
given by
\begin{align}
\nonumber
g_t^H&=\frac{\sin\alpha}{\sin\beta}=-\sba\frac{1}{\tb}+\cba,\\
\text{Type I: } g_b^H&=\frac{\sin\alpha}{\sin\beta}=-\sba\frac{1}{\tb}+\cba,\quad
\text{Type II: } g_b^H=\frac{\cos\alpha}{\cos\beta}=\sba\tb+\cba\quad.
\label{eq:gtbH}
\end{align}

In the decoupling limit, $|\sba| \to 1$, the light Higgs boson $h$ couples 
to the gauge bosons with the 
same strength as the \sm{} Higgs boson. 
In contrast, the more \sm-like the coupling of the light Higgs boson to
gauge bosons is, the more suppressed is the coupling of the heavy
Higgs boson~$g_V^H$ to gauge bosons, as a consequence of the well-known 
sum rule $(g_V^h)^2+(g_V^H)^2=1$.
Accordingly, for a \thdm{} with a \sm-like Higgs boson $h$ at $125$\,GeV the 
signal for the production of the heavier Higgs~$H$
in $gg\rightarrow H\rightarrow VV$ will necessarily be rather weak.

For a suppressed signal of this kind interference effects can be important,
both with the off-shell light Higgs boson contribution
as well as with the background diagrams. Our analysis will quantify
these effects for different final states as a function of $\sba$,
$\tb$ as well as the employed Higgs mass $\mH$. 
In the context of interference effects obviously also the 
width, $\GaH$, is an important quantity. In our analysis $\GaH$ is not
treated as a free parameter, but it is calculated from the other \thdm\ parameters
with the help of {\tt 2HDMC}.

\subsection{Details of the calculation}
\label{sec:detailscalc}

\begin{figure}[htp]
\begin{center}
\begin{tabular}{ccc}
\includegraphics[width=0.31\textwidth]{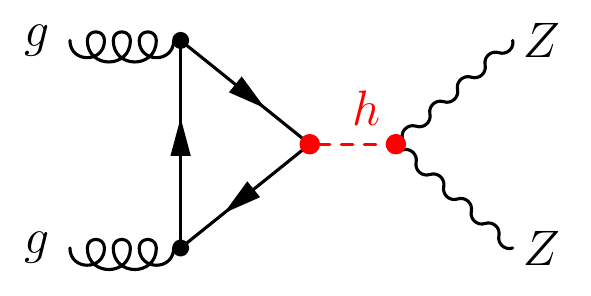} & 
\includegraphics[width=0.31\textwidth]{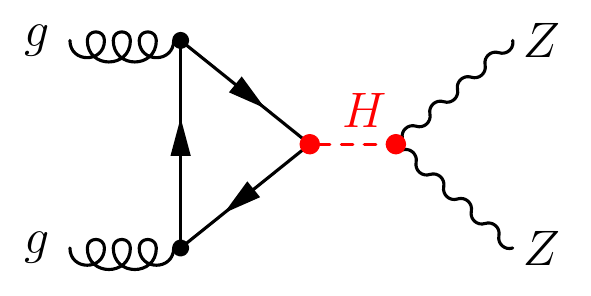} & 
\includegraphics[width=0.31\textwidth]{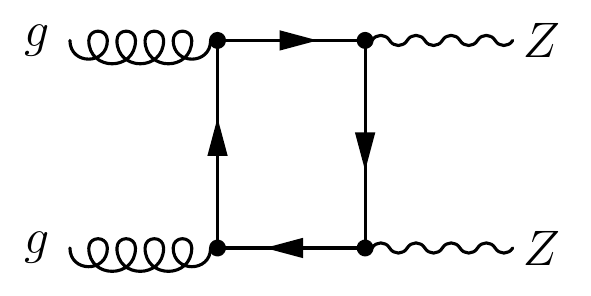} \\[-0.2cm]
 (a) & (b) & (c)
\end{tabular}
\end{center}
\vspace{-0.6cm}
\caption{Feynman diagrams for $gg\rightarrow ZZ$.}
\label{fig:feyngg2ZZ} 
\end{figure}

We briefly describe here the calculations that we have carried out for
the process with two on-shell $Z$ bosons, $gg\rightarrow ZZ$, and for the
full process with four fermions in the final state,
$gg\rightarrow e^{+}e^{-}\mu^{+}\mu^{-}$, 
$gg\rightarrow e^{+}e^{-}\nu_{\mu}\bar{\nu}_{\mu}$ and 
$gg\rightarrow e^{+}e^{-}\nu_e\bar{\nu}_e$.
Our implementation of $gg\rightarrow ZZ$ within {\tt vh@nnlo} was
generated with the help of {\tt FeynArts}~\cite{Hahn:2000kx}
and {\tt FormCalc}~\cite{Hahn:1998yk}, see the diagrams in 
\fig{fig:feyngg2ZZ}. The implementation links to
{\tt LoopTools}~\cite{Hahn:1998yk} for the calculation of the employed one-loop 
Feynman integrals and to {\tt 2HDMC}~\cite{Eriksson:2009ws} for the
calculation of the Higgs boson widths~$\Gamma_h$ and $\GaH$.
In comparison to our treatment of the full process, see below, we
employ additional approximations for the process 
$gg\rightarrow ZZ$, for which we investigate the different interference
contributions. In particular, we do not take into account
the top-quark contribution to the box diagrams in 
$gg\rightarrow ZZ$,
see \fig{fig:feyngg2ZZ}~(c), even though they are of relevance for large invariant masses of the gauge bosons above $2m_t$.
However, the top-quark contribution to those diagrams does not add new features to our qualitative discussion
of interference effects in \sct{sec:ggZZ}.
We treat the remaining five quarks massless in contrast to
our {\tt GoSam} implementation, which includes all six quarks with finite top- and bottom-quark masses, see below.
The triangle diagrams with an intermediate light or heavy Higgs boson, see \fig{fig:feyngg2ZZ}~(a) and (b), take into
account the massive top-quark and bottom-quark contributions.

\begin{figure}[htp]
\begin{center}
\begin{tabular}{ccc}
\includegraphics[width=0.31\textwidth]{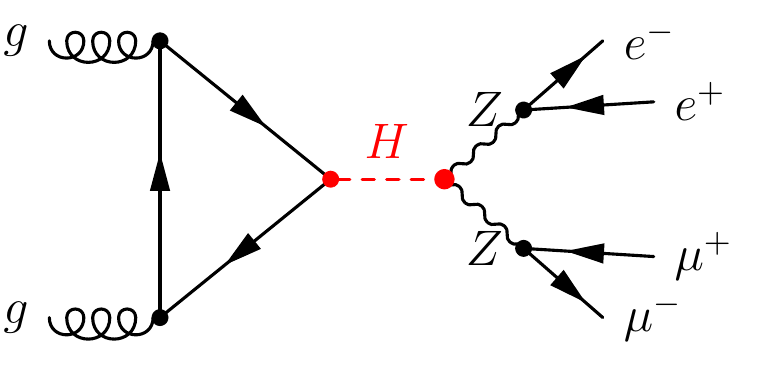} & 
\includegraphics[width=0.31\textwidth]{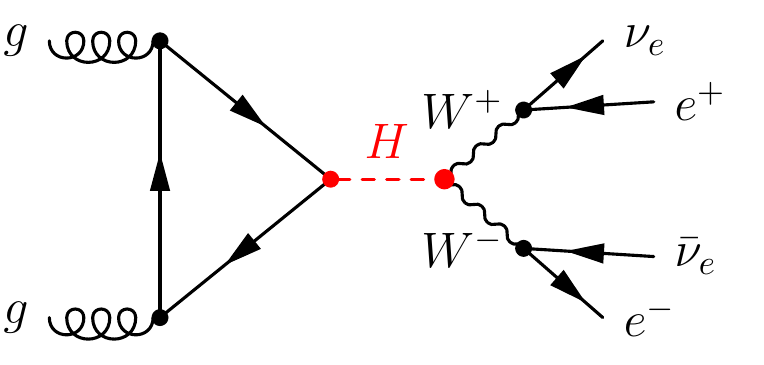} & 
\includegraphics[width=0.31\textwidth]{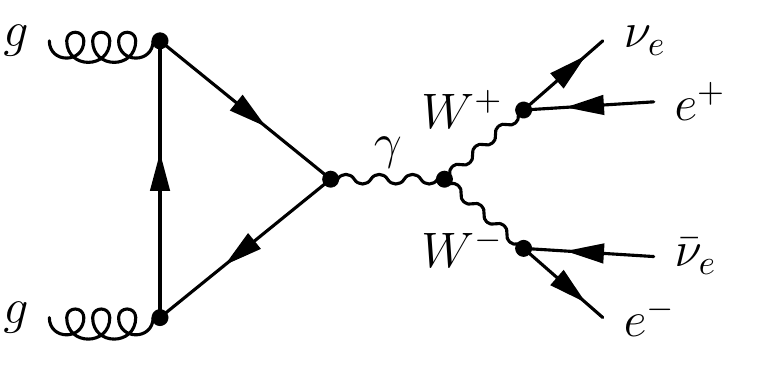} \\[-0.2cm]
 (a) & (b) & (c)
\end{tabular}
\begin{tabular}{cc}
\includegraphics[width=0.3\textwidth]{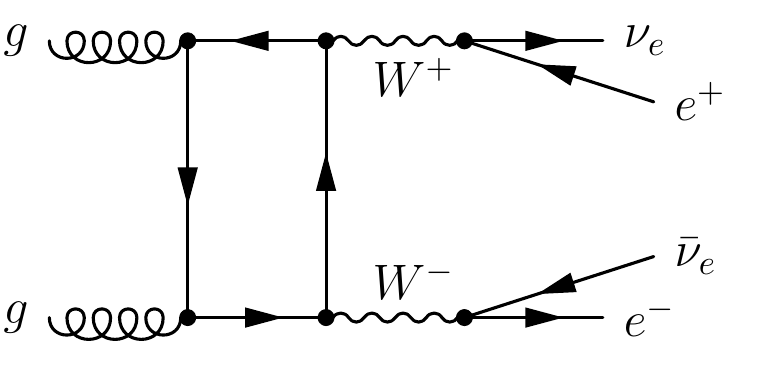} &
\includegraphics[width=0.3\textwidth]{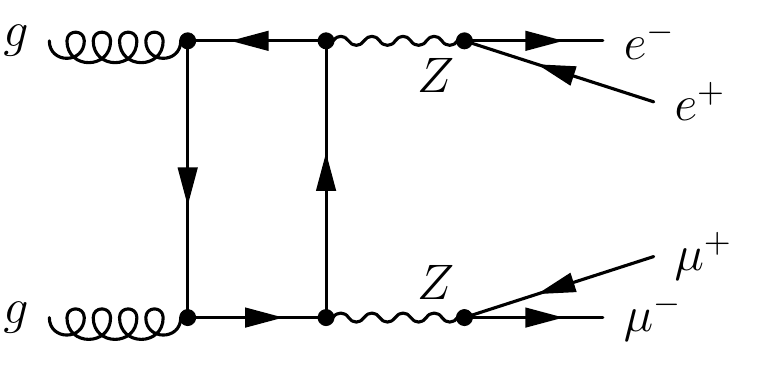} \\[-0.2cm]
 (d) & (e)
\end{tabular}
\end{center}
\vspace{-0.6cm}
\caption{Sample Feynman diagrams for double resonant $W,Z$ contributions, i.e. where the final state leptons directly come from
  the decay of massive gauge bosons.}
\label{fig:diags_res} 
\end{figure}

The amplitudes for the processes $gg\rightarrow e^{+}e^{-}\mu^{+}\mu^{-}$, 
$gg\rightarrow e^{+}e^{-}\nu_{\mu}\bar{\nu}_{\mu}$ and $gg\rightarrow e^{+}e^{-}\nu_e\bar{\nu}_e$
have been generated with {\tt GoSam}~\cite{Cullen:2011ac,Cullen:2014yla}. {\tt GoSam} is a publicly
available tool for the automated generation of one-loop amplitudes within and beyond the Standard Model.
It is based on a Feynman diagrammatic approach, where the Feynman diagrams are first generated with
{\tt QGraf}~\cite{Nogueira:1991ex} and
{\tt Form}~\cite{Vermaseren:2000nd,Kuipers:2012rf}, and
{\tt Spinney}~\cite{Cullen:2010jv}, {\tt Haggies}~\cite{Reiter:2009ts} and
{\tt Form} are used to write an optimized Fortran output. For the reduction of the tensor integrals 
there are several options available.
We used {\tt Ninja}~\cite{Mastrolia:2012bu,vanDeurzen:2013saa,Peraro:2014cba}, an automated package 
for integrand reduction via Laurent expansion.
Alternatively one can use other reduction
techniques such as integrand reduction in the OPP method~\cite{Ossola:2006us,Mastrolia:2008jb,Ossola:2008xq} as implemented in
{\tt Samurai}~\cite{Mastrolia:2010nb} or methods of tensor integral reduction as implemented in
{\tt Golem95}~\cite{Heinrich:2010ax,Binoth:2008uq,Cullen:2011kv,Guillet:2013msa}.
The resulting scalar integrals are evaluated using {\tt OneLOop}~\cite{vanHameren:2010cp}.\\
In this case the implementation of a \thdm{} model in {\tt GoSam} requires only the implementation of a second Higgs boson, while
leaving the relative couplings $g_f^{h,H}$ and $g_V^{h,H}$ as free parameters, which can be modified according
to the specific parameters that are considered.
Our discussion including the different decay channels of the intermediate
vector bosons considers final states with four leptons, namely
\begin{equation}
 gg \to e^{+}e^{-}\mu^{+}\mu^{-}, \quad gg \to e^{+}e^{-}\nu_{\mu/\tau}\bar{\nu}_{\mu/\tau}, \quad
 gg \to e^{+}e^{-}\nu_{e}\bar{\nu}_{e}\;,
 \label{eq:subprocesses}
\end{equation}
where we have to sum over all possible intermediate configurations leading to the given final state.
This particularly means that depending on the sub-process, also intermediate $W$ bosons as well as non-resonant
contributions and photon exchange have to be taken into account. In \fig{fig:diags_res} we show a few
sample diagrams that contribute to a resonant decay of the massive gauge bosons, either for the actual signal
(including a Higgs boson) or to the background (without an intermediate Higgs boson exchange).

\begin{figure}[htp]
\begin{center}
\begin{tabular}{ccc}
\includegraphics[width=0.31\textwidth]{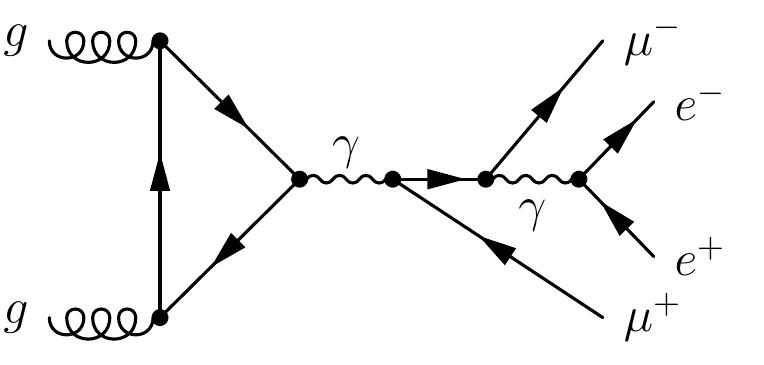} & 
\includegraphics[width=0.31\textwidth]{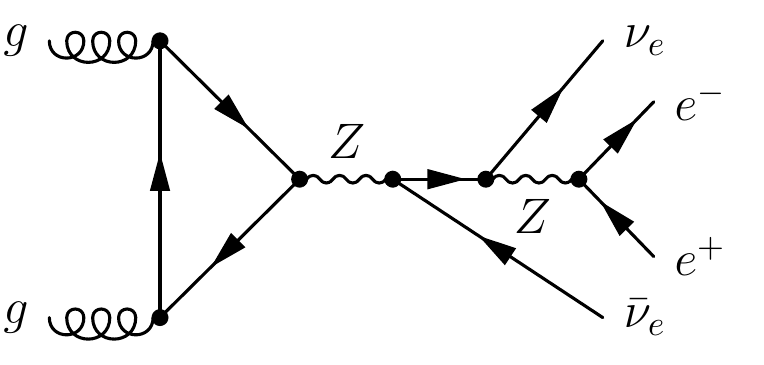} & 
\includegraphics[width=0.31\textwidth]{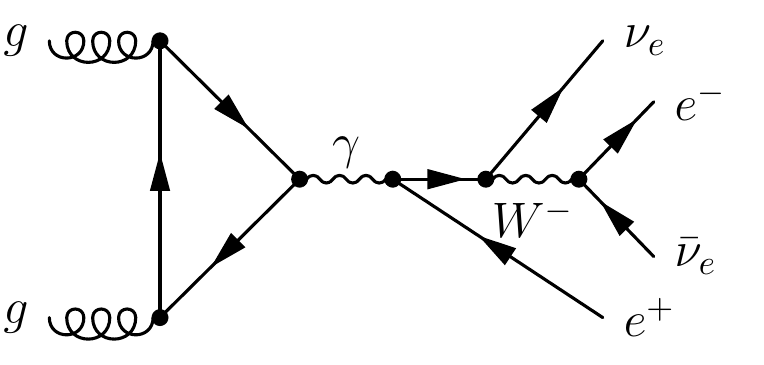} \\[-0.2cm]
 (a) & (b) & (c)
\end{tabular}
\end{center}
\vspace{-0.6cm}
\caption{Sample Feynman diagrams for single- and non-resonant $W,Z$ contributions for the sub-processes under consideration.}
\label{fig:diags_nonres} 
\end{figure}

In \fig{fig:diags_nonres} we show a few sample diagrams for single- or non-resonant $W,Z$ contributions to the 
same four lepton final states. These types of contributions make
it necessary to impose certain cuts 
on the final state leptons to render the cross section finite, see below.

For the box diagrams, see e.g. \fig{fig:diags_res}~(d) and (e),
we take into account all six quark flavors, where the first four are considered to be massless. For diagrams
involving a Higgs coupling to quarks only top and bottom quarks are of relevance, since
lighter quark contributions are suppressed by their small Yukawa couplings.
For the numerical integration over the four particle phase space we have combined the {\tt GoSam} amplitudes
with the integration routines provided by {\tt MadEvent}~\cite{Maltoni:2002qb,Alwall:2007st}.

It is well-known that the calculation of processes including internal Higgs
bosons, in particular if one includes higher orders,
needs a gauge invariant formulation of the Higgs boson
propagator. Since we are working at \lo{} \qcd{} only,
a simplistic Breit-Wigner propagator is sufficient for all our purposes.

We checked our modified {\tt vh@nnlo} and our {\tt GoSam} implementations
against each other for $gg\rightarrow ZZ$ at the amplitude level 
and reproduced parts of the results presented in \citere{Kauer:2015hia} for
the four leptonic final state within the numerical uncertainties.

While for the purpose of brevity of our presentation we do not
give explicit analytical results for the final states with four fermions,
we refer the reader to \citere{Glover:1988rg} for the helicity amplitudes
of $gg\to ZZ$ (taking into account the correction described in \citere{Jung:2015sna}).
The amplitudes given for the \sm{} can individually be translated to the \thdm{} case
by multiplying them with effective couplings of Higgs bosons to fermions and gauge
bosons. The split in longitudinally and transversely polarized final
state $Z$ bosons is instructive, since the relative fraction of
longitudinal $Z$~bosons rises with increasing invariant
masses of the two $Z$~bosons. This rise also explains the relevance of the
top-quark contribution for the background amplitudes
involving longitudinal $Z$ bosons, since they are proportional to $m_t^2$.
We focus in the following on our numerical analysis
in order to discuss the dependences on all relevant \thdm{} parameters.
For a discussion of some of the qualitative features in terms of contributions 
from real and imaginary parts to the interferences
we also refer to \citere{Jung:2015sna}.

In our numerical analysis below we will first study the full process
with four fermions in the final state 
generated with {\tt GoSam} in order to discuss the experimental
sensitivities in terms of the relevant distributions. In a second step we
will focus on the on-shell approximation for two $Z$ bosons, i.e.\ the
simplified process $gg\rightarrow ZZ$, 
in order to quantify the relevance of interference contributions.

\section{Parameter choice and selection cuts}
\label{sec:paramchoice}

We consider five benchmark scenarios to cover different aspects of a heavy Higgs boson
in the phenomenology of a \thdm{}.
The scenarios yield sufficiently high event rates for the heavy Higgs boson
to be potentially observable at the \lhc{} at least at very high integrated luminosities. We also vary
certain parameters of the scenarios in order to discuss the relevance of the interference of
the heavy Higgs signal with the light Higgs boson and the background.
All scenarios include a light Higgs boson with mass $m_h=125$\,GeV.
We keep the couplings of the light Higgs close to the ones of the \sm{} Higgs
by a proper choice of $\tb$ and $\sba$. The masses (and widths) of quarks
and gauge bosons are set to
\begin{align}\nonumber
 &m_t=172.3\,\text{GeV}, &m_b(m_b)=4.16\,\text{GeV},\\\nonumber
 &\mz=91.1876\,\text{GeV}, &\mw=80.398\,\text{GeV},\\
 &\Gaz = 2.4952\,\text{GeV}, &\Gaw= 2.085\,\text{GeV}.
\end{align}
To keep our calculation simple,
we work with the $\overline{\text{MS}}$ bottom-quark mass as input to the bottom-Yukawa coupling
as well as to internal propagators.
Note that in all cases we choose the masses of the pseudoscalar $A$ and the charged Higgs boson $H^\pm$ heavy enough not
to open decay modes of the heavy Higgs boson $H$ into them.
The corresponding width of the heavy Higgs boson~$\GaH$ is obtained with {\tt 2HDMC}.
The detailed settings of the scenarios are presented in \tab{tab:2hdm}.

\begin{table}[htb]
\begin{center}
\begin{tabular}{| c || c | c | c | c | c |}
\hline
Scenario & \thdm{} type &$\tb$ & $\sba$ & $\mH$ & $\GaH$
\\\hline\hline
S1 & II & $2$   & $-0.995$  & $200$\,GeV & $0.0277$\,GeV\\\hline
S2 & II & $1$   & $0.990$   & $400$\,GeV & $3.605$\,GeV\\\hline
S3 & I  & $5$   & $0.950$   & $400$\,GeV & $2.541$\,GeV\\\hline
S4 & I  & $5$   & $0.96695$ & $200$\,GeV & $0.0882$\,GeV\\\hline
S5 & II & $20$  & $0.990$   & $400$\,GeV & $5.120$\,GeV\\\hline
\hline
\end{tabular}
\end{center}
\vspace{-5mm}
\caption{\thdm{} scenarios considered in our analysis.}
\label{tab:2hdm}
\end{table}

Scenario~S1 is a standard \thdm{} scenario of type II, similarly to S2. Both
have a large value of $|\sba|$ close to $1$, such that the
coupling of the heavy Higgs to gauge bosons proportional to $\cba$ is small.
They in particular differ in the choice of the heavy Higgs mass $\mH$.
Scenario~S1 is inspired by the ATLAS analysis carried out in \citere{Aad:2015kna}, where we
want to discuss the relevance of interferences for the performed experimental searches
in the previous \lhc{} run at $\sqrt{s}=8$\,TeV.

Scenario~S3 is of type I, where larger values of $\cba$ are
still compatible with data, see \citere{Haber:2015pua}.
The last two scenarios~S4 and S5 are such that particularly large 
interferences are possible either between the heavy Higgs boson signal and the light Higgs boson
or the background. Potentially large interference effects can occur for values 
of $\sba$ where the top- (and bottom-quark) Yukawa coupling $g_t^H$ (and $g_b^H$)
for the heavy Higgs boson in a \thdm{} of type~II (or I) are
suppressed. Another possibility is a relatively large value of $\tb$, which
increases the relevance of the bottom-quark Yukawa coupling $g_b^H$ for the heavy Higgs boson
in a \thdm{} of type~II.
A general difference between scenarios~S2, S3 and S5 and scenarios S1 and S4
is also the heavy Higgs boson mass. The latter is once above and once
below the top threshold $2m_t$.
For $\mH>2m_t$ on the one hand the decay mode $H\rightarrow t\bar t$ opens and on the other hand
also an imaginary part of the amplitude $gg\rightarrow H\rightarrow VV$ is induced
through the top-quark loop, which
is of importance for interferences, see also \citere{Jung:2015sna}. 

For completeness we also investigated a ``flipped Yukawa''
scenario, see e.g. \citeres{Ferreira:2014naa,Ferreira:2014qda,Haber:2015pua},
where the relative bottom Yukawa coupling of the light Higgs is $g_b^h=-1$.
Such a scenario makes it possible to have a relatively
large value for $\cba$ and thus a large
coupling of the heavy Higgs boson to gauge bosons keeping the light Higgs compatible
with experimental bounds. However, in the processes under consideration
it does not provide new features
with respect to the other five scenarios and is thus not listed separately.

Our studies are generally carried out for the \lhc{} with a centre-of-mass energy of $\sqrt{s}=13$\,TeV,
except for scenario~S1 which is investigated both at $8$\,TeV and $13$\,TeV.
The role of interference effects is a bit less pronounced at $7/8$\, TeV compared to  $13$\,TeV. 
We make use of {\tt CT10nnlo}~\cite{Gao:2013xoa} as PDF set for the gluon luminosities.
Since our calculations are purely performed at \lo{}, the renormalization scale dependence enters
through the strong coupling~$\alpha_s$ only, which we take from the employed PDF set.
We choose the renormalization and factorization scale to be dynamical, namely
half of the invariant mass of the gauge boson system $\mu_R=\mu_F=m_{VV}/2$, i.e. $\mu_R=\mu_F=m_{4l}/2$
in case of the four leptonic final states.
Since the Gram determinants of the box diagrams, see \fig{fig:feyngg2ZZ}~(c), approach zero for low $p_T$ of the heavy
gauge bosons, we perform a technical cut of $p_T^Z>2$\,GeV and $p_T^W>2$\,GeV for all our processes. 
It is known to have a small effect on the cross section~\cite{Kauer:2012ma,Campbell:2013una},
which we have numerically confirmed for the processes under consideration.

For the processes with four charged leptons or two charged leptons and two neutrinos in the final
state, we additionally cut on
the transverse momentum and the pseudorapidity of each lepton~$l$, $p_T^l>10$\,GeV and $|\eta_l|<2.7$,
the $R$-separation between individual leptons $R^{ll'}>0.1$, as well as $m_{ll}>5$\,GeV, where
$ll$ is an oppositely charged same-flavour dilepton pair.
For the neutrinos we ask for a total missing transverse momentum of $E_T^{\text{miss}}>70$\,GeV.
The cuts are inspired by the recent ATLAS analysis carried out in \citere{Aad:2015kna}.

One of the most important observables is certainly the invariant mass
distribution of the four leptons, as the two Higgs 
bosons manifest themselves in Breit-Wigner peaks in this distribution.
For the first process $gg\rightarrow e^+e^-\mu^+\mu^-$ of \eqn{eq:subprocesses} this observable $m_{4l}$ is also experimentally
easily accessible due to two electrons and two muons in the final state. In the cases with neutrinos in the final state 
the situation is more involved. The invariant mass is no longer an observable that is experimentally accessible but only
a transverse component can be measured. 
As we are particularly interested in a heavy Higgs boson that will decay into the four leptons via 
two intermediate electroweak gauge bosons, a sensible choice is to consider the transverse mass of the underlying two boson
system. In our case the two boson system can be $ZZ$ as well as $WW$. 
We therefore define a general transverse mass via \cite{Aad:2015agg}
\begin{equation}
 m_{VV,T}^2= \left(E_{T,ll} +E_{T,\nu\nu}\right)^2 - \left|\vec{p}_{T,ll} + \vec{p}_{T,\nu\nu}\right|^2\;,
 \label{eq:mT}
\end{equation}
with
\begin{equation}
 E_{T,ll}=\sqrt{p_{ll}^2+|\vec{p}_{T,ll}|^2}\;,\quad \text{and}\;\; E_T^{\text{miss}}=E_{T,\nu\nu}=\left|\vec{p}_{T,\nu\nu}\right|\;.
 \label{eq:ET}
\end{equation}
In the case of a purely leptonic final state the last squared term in \eqn{eq:mT} vanishes.
As we are interested in the heavy Higgs boson and its interference with the light Higgs boson and the
background we put an additional cut on the invariant mass.
For the scenarios where the heavy Higgs mass is $400$\,GeV, we require
$m_{4l}>350$\,GeV for the muonic process. For the neutrino process we 
apply the same cut but on $m_{VV,T}$. For the scenarios where the heavy Higgs boson mass
is $200$\,GeV, we choose the invariant mass cut as $m_{4l}>100$\,GeV or $m_{VV,T}>100$\,GeV, respectively.

\section{Numerical results}
\label{sec:results}

We present our numerical results as follows: We start with a discussion of the four fermionic final states
making use of the benchmarks scenarios defined in \sct{sec:paramchoice}.
Afterwards we exemplify the relevance of interference effects for $gg\rightarrow ZZ$, where
we vary either the heavy Higgs mass $\mH$,
the relative coupling to gauge bosons $\sba$ ($\cba$ for the heavy Higgs respectively) or the ratio of vacuum
expectation values $\tb$ and fix the other parameters according to the benchmark scenarios. The three
mentioned free parameters are the ones relevant for the phenomenology of the heavy Higgs boson
in the processes under consideration. As noted the heavy Higgs width~$\GaH$ is obtained from {\tt 2HDMC}.

The process~$gg\rightarrow VV(\rightarrow e^+e^-\mu^+\mu^-/e^+e^-\nu_l\bar\nu_l)$ is by far superseded
by the large background $q\bar q\rightarrow VV(\rightarrow e^+e^-\mu^+\mu^-/e^+e^-\nu_l\bar\nu_l)$ (without Higgs boson contributions),
in particular at high invariant masses.
However the two processes do not interfere and can be added incoherently.
For the relevance of other backgrounds we refer to \citere{Aad:2015kna}.
A detailed simulation with all backgrounds in order to determine
to which cross sections the experiments are actually sensitive at high integrated
luminosities is beyond the scope of this paper. In addition,
with increasing $\tb$ in a \thdm{} of
type II the bottom-quark initiated processes
$b\bar b\rightarrow VV(\rightarrow e^+e^-\mu^+\mu^-/e^+e^-\nu_l\bar\nu_l)$ 
become relevant, which involve Higgs boson contributions
and thus similar interferences as observed in the gluon initiated processes.

\subsection{Discussion of four-fermion final states}
\label{sec:fourfermionic}

In the following we discuss the processes
\begin{equation}
 gg \to e^{+}e^{-}\mu^{+}\mu^{-}, \quad gg \to e^{+}e^{-}\nu_{\mu/\tau}\bar{\nu}_{\mu/\tau}, \quad
 gg \to e^{+}e^{-}\nu_{e}\bar{\nu}_{e}\;,
\end{equation}
for the benchmark scenarios and the cuts shown in \sct{sec:paramchoice},
taking into account the contributions displayed in \sct{sec:detailscalc}.

\begin{figure}[htb]
\begin{center}
\begin{tabular}{cc}
\includegraphics[width=0.5\textwidth]{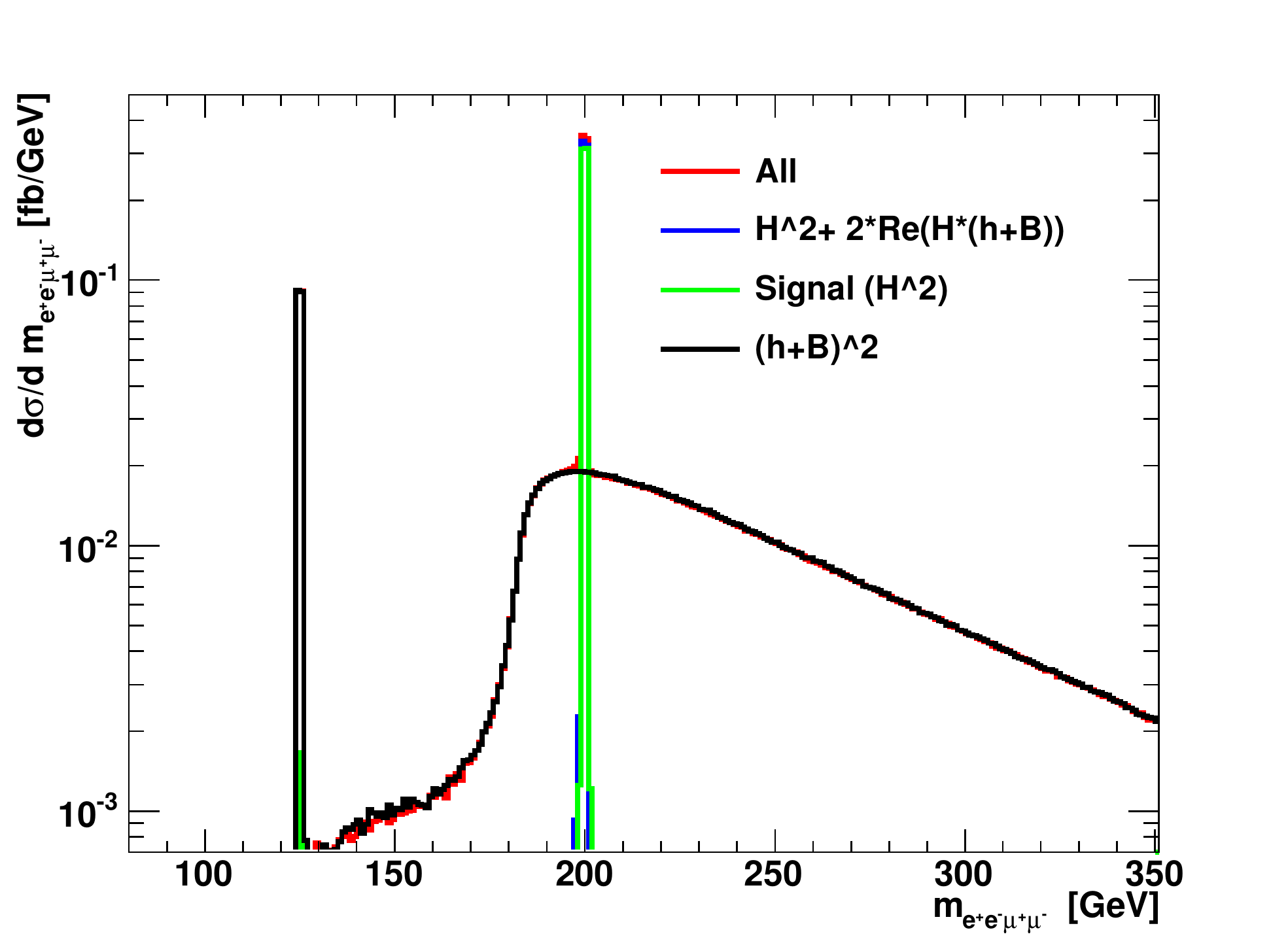} &
\includegraphics[width=0.5\textwidth]{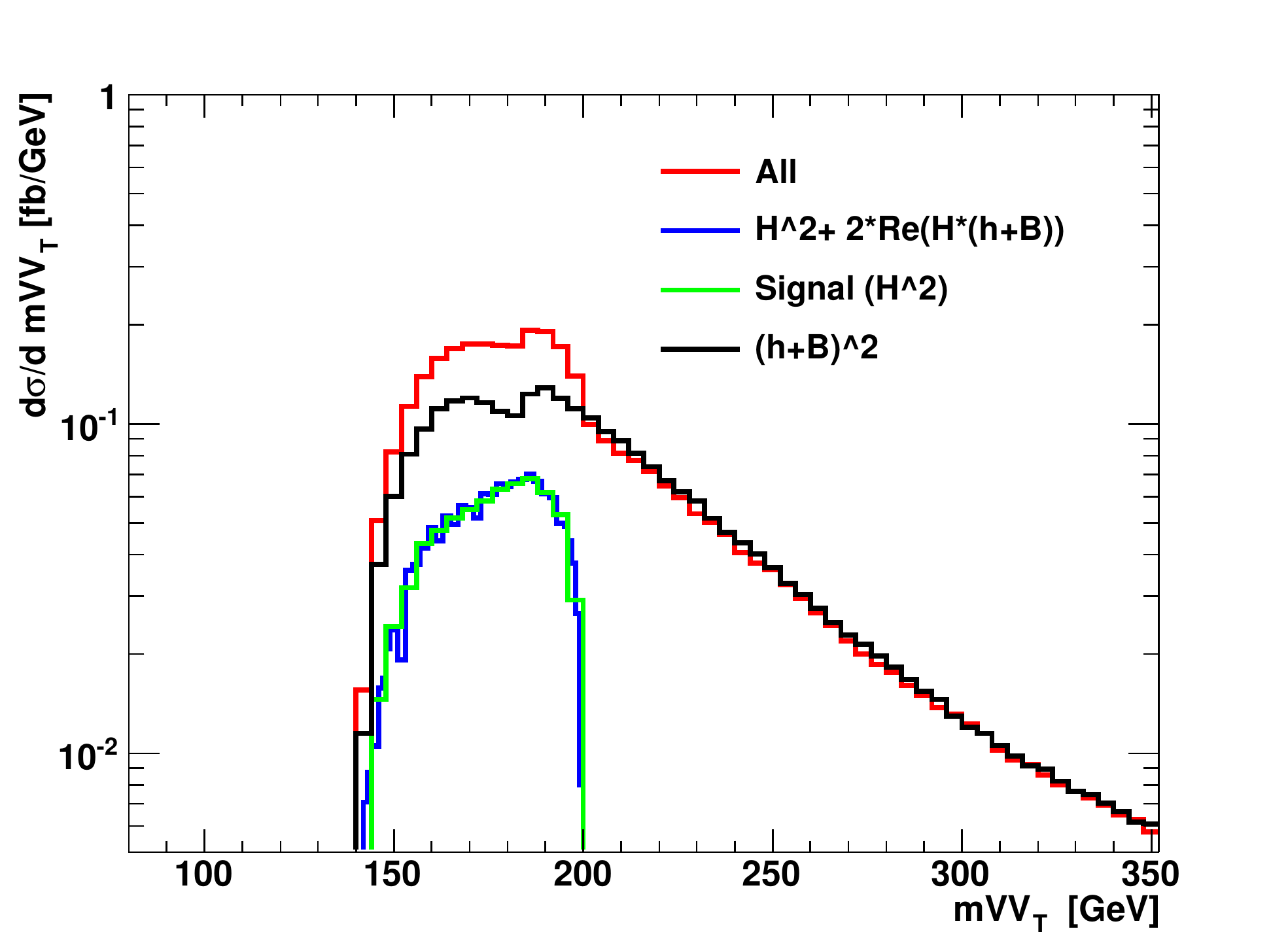}\\[-0.2cm]
 (a) & (b)
\end{tabular}
\end{center}
\vspace{-0.6cm}
\caption{(a) Invariant mass distribution for $gg\rightarrow e^{+}e^{-}\mu^{+}\mu^{-}$ and (b) transverse mass
distribution for $gg\rightarrow e^{+}e^{-}\nu_l\bar\nu_l$
for scenario~S1 at $\sqrt{s}=13$\,TeV.}
\label{fig:S1_mass} 
\end{figure}

We start the discussion of the numerical results with scenario~S1.
\fig{fig:S1_mass} shows the invariant mass distribution of the four leptons for $gg\rightarrow e^{+}e^{-}\mu^{+}\mu^{-}$ and 
the transverse mass distribution using the definition in \eqn{eq:mT} for the processes involving final state neutrinos. 
In this plot and in the following we distinguish four different contributions. In red, denoted with ``All'', we plot all
contributions that lead to the given final state in the considered scenario. In green, we only plot the contribution from
the heavy Higgs boson, whereas in blue we also add the interference of the heavy Higgs boson with the background and the light Higgs boson.
The contribution $|h+B|^2$, plotted in black, contains besides the contributions without any Higgs also contributions of the light Higgs as well as the interference
contributions of the light Higgs boson with non-Higgs diagrams.\\
In the invariant mass plot of $gg\rightarrow e^{+}e^{-}\mu^{+}\mu^{-}$, see \fig{fig:S1_mass}~(a),
the two Higgs boson peaks at $m_{4l}=125$ and $200$\,GeV can be clearly seen.
Due to the very small width of the heavy Higgs boson there is no distortion of the Breit-Wigner shape visible, and also the impact
of the interference contribution to the total height of the peak is rather small. The transverse mass distribution 
for $gg\rightarrow e^{+}e^{-}\nu_l\bar\nu_l$, see \fig{fig:S1_mass}~(b), shows a quite different pattern. First of all there is no peak from the light Higgs boson.
The reason for this are the different cuts compared to the process without neutrinos. The requirement of $E_T^{\text{miss}}>70$\,GeV
excludes this region of phase space. 
Due to the fact that the four-momenta of the neutrinos are experimentally not accessible one sets
$E_{T,\nu\nu}=\left|\vec{p}_{T,\nu\nu}\right|$, which ignores the invariant mass of the neutrino system.
This removes the sharp peak of the heavy Higgs boson, 
which is visible in the invariant mass distribution of the muon process.
Instead of a distinguished peak one obtains a broad distribution.
But also here the contribution of the interference remains small. 
A second difference compared to the muon process is the occurrence of a small dip at around $m_{VV,T}=180$\,GeV in both signal and background.
This specific shape is due to the fact that the total contribution to the process with neutrino final state consists
of the sum of two different sub-processes, namely the one with the electron neutrino and the ones with muon- and tau neutrino
in the final state.
Whereas the first sub-process also has contributions from intermediate $W$-bosons, this is not the case for the latter sub-processes.
The two sub-processes therefore show a different kinematical behaviour, and the sum of the two contributions leads to the given
distribution.

\begin{figure}[h!]
%\begin{figure}[htb]
\vspace{-0.4cm}
\begin{center}
\begin{tabular}{cc}
\includegraphics[width=0.5\textwidth]{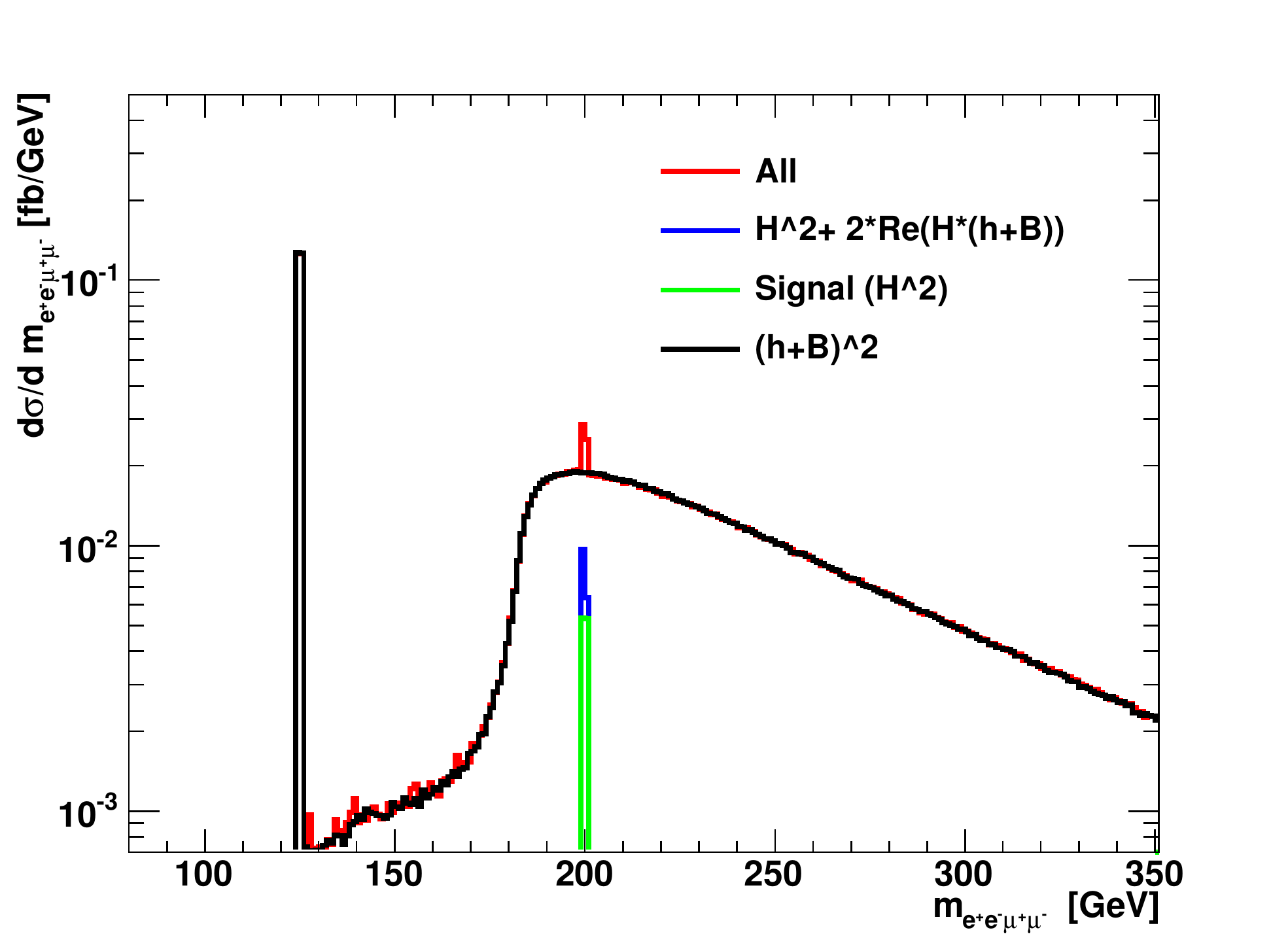} &
\includegraphics[width=0.5\textwidth]{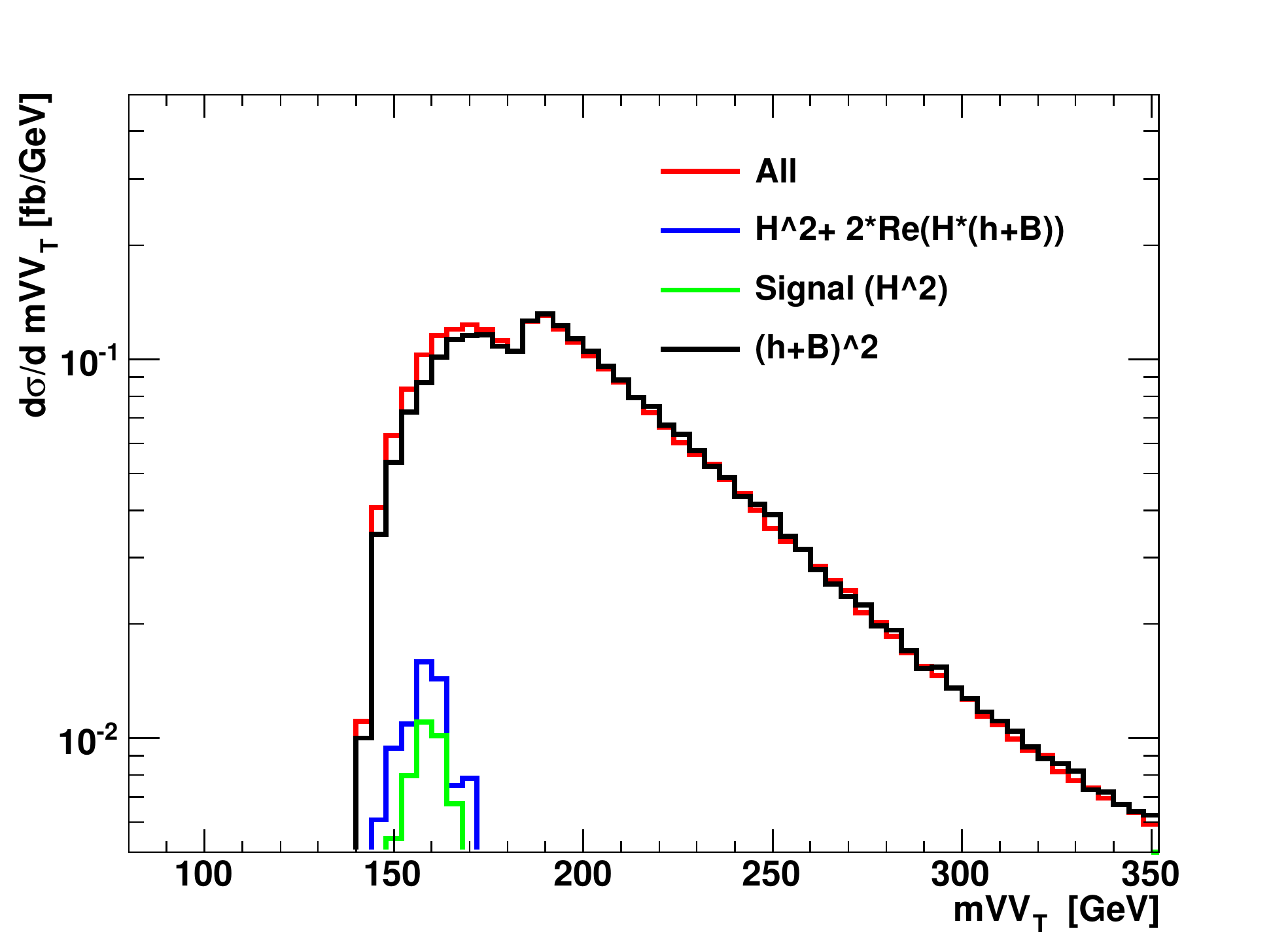}\\[-0.2cm]
 (a) & (b)
\end{tabular}
\end{center}
\vspace{-0.6cm}
\caption{(a) Invariant mass distribution for $gg\rightarrow e^{+}e^{-}\mu^{+}\mu^{-}$ and (b) transverse mass
distribution for $gg\rightarrow e^{+}e^{-}\nu_l\bar\nu_l$
for scenario~S4 at $\sqrt{s}=13$\,TeV.}
\label{fig:S4_mass} 
%\end{figure}
%\begin{figure}[htp]
\vspace{-0.4cm}
\begin{center}
\begin{tabular}{cc}
\includegraphics[width=0.5\textwidth]{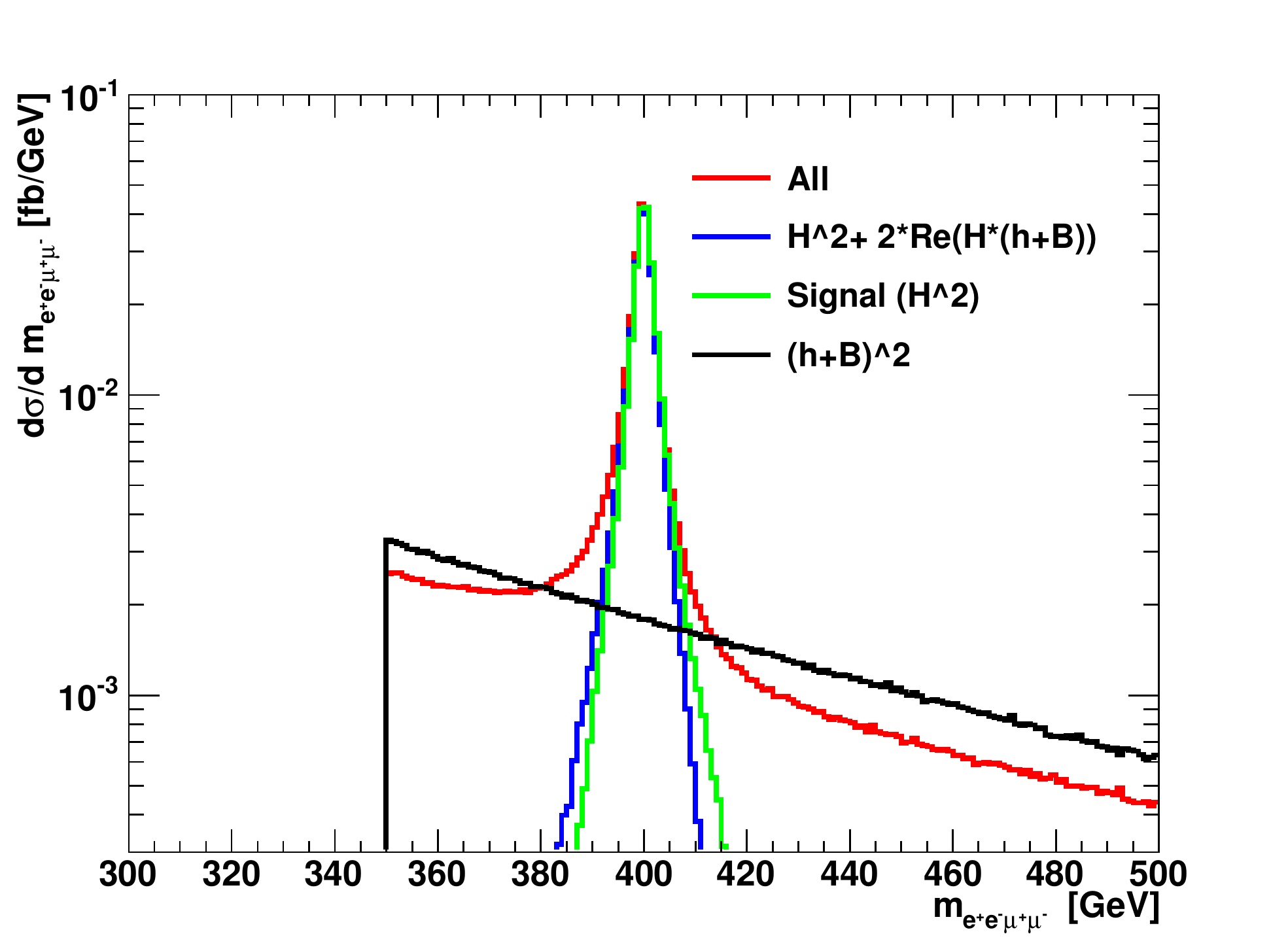} &
\includegraphics[width=0.5\textwidth]{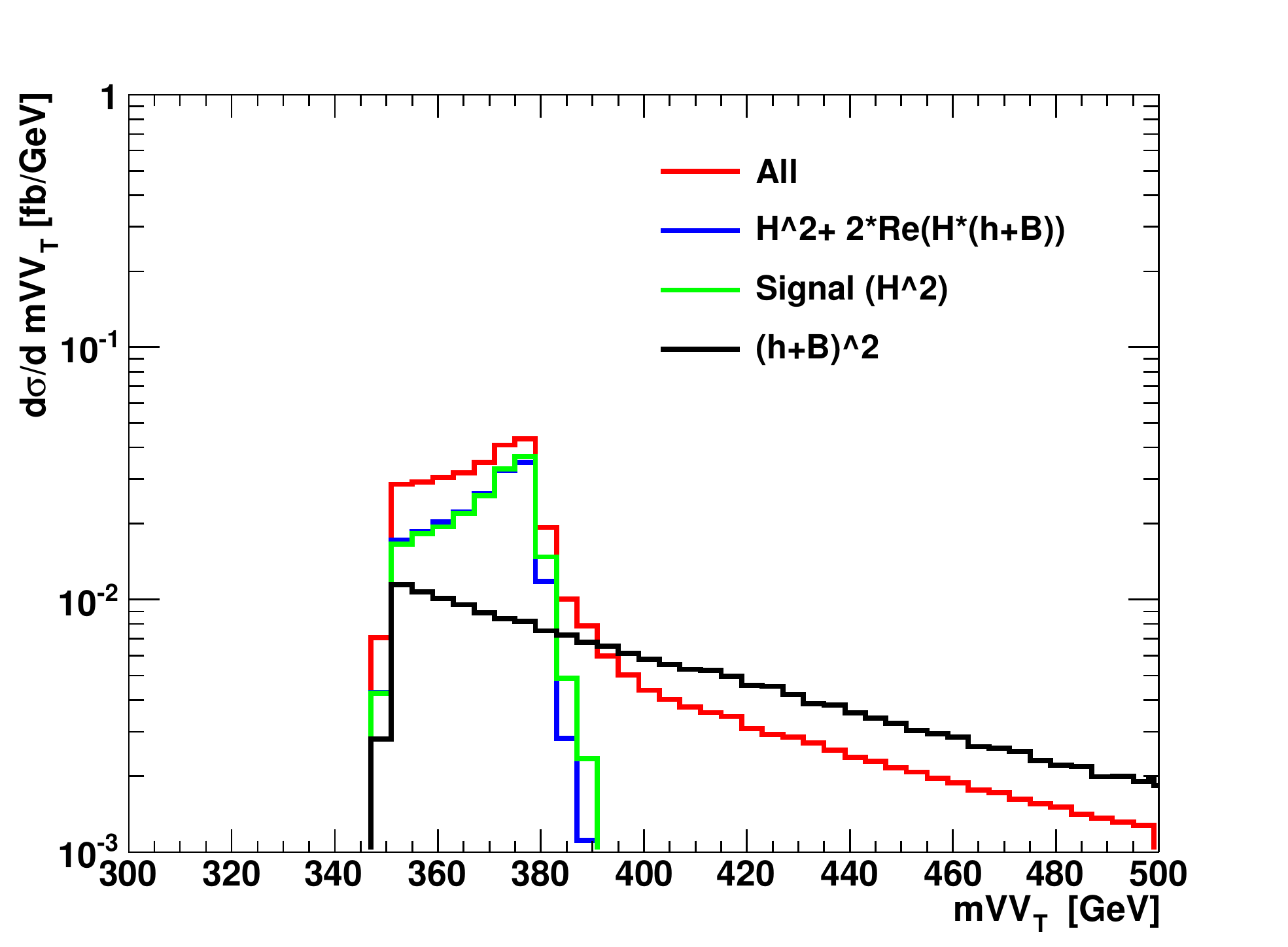}\\[-0.2cm]
 (a) & (b)
\end{tabular}
\end{center}
\vspace{-0.6cm}
\caption{(a) Invariant mass distribution for $gg\rightarrow e^{+}e^{-}\mu^{+}\mu^{-}$ and (b) transverse mass
distribution for $gg\rightarrow e^{+}e^{-}\nu_l\bar\nu_l$
for scenario~S2 at $\sqrt{s}=13$\,TeV.}
\label{fig:S2_mass} 
\end{figure}

Scenario~S4 shown in \fig{fig:S4_mass} is phenomenologically similar to scenario~S1 with a very narrow heavy Higgs boson peak.
In contrast to scenario~S1 the couplings of the heavy Higgs to the fermions and gauge bosons are smaller leading to a small
heavy Higgs boson signal. 
The relative importance of the interference effect is sizeable and increases
the heavy Higgs boson signal by roughly a factor of $2$. However its measurement
remains challenging. For the neutrino process on the r.h.s. one observes a more pronounced dip as compared to scenario S1.
In contrast to the type~I scenario~S4, the type~II scenario~S2 leads to a large signal-over-background ratio for the heavy Higgs peak as can 
be seen from \fig{fig:S2_mass}. Accordingly the interference contribution leads only to a mild distortion of the shape and 
thus only has a small impact.

In general, comparing $gg\rightarrow e^+e^-\mu^+\mu^-$ with $gg\rightarrow e^+e^-\nu_l\bar\nu_l$,
the big advantage of the latter process is the much larger total cross section.
We find that for $\sqrt{s}=13$\,TeV the cross section for the process with
neutrinos in the final state is roughly one order of magnitude
larger than for $gg\rightarrow e^+e^-\mu^+\mu^-$. First there are three neutrino generations involved,
second there are more additional underlying sub-processes (including internal $W$ bosons) and different
couplings. On the other hand $gg\rightarrow e^{+}e^{-}\nu_l\bar\nu_l$ is experimentally more difficult to access than 
$gg\rightarrow e^{+}e^{-}\mu^{+}\mu^{-}$ with four tagged leptons in two different flavors.
For the latter process mass windows around the $Z$-boson mass even allow to cut away non-resonant contributions
and only measure the pure $gg\to ZZ$ contribution.

\setlength{\tabcolsep}{0pt}
\begin{figure}[h!]
\begin{center}
\begin{tabular}{cc}
\includegraphics[width=0.5\textwidth]{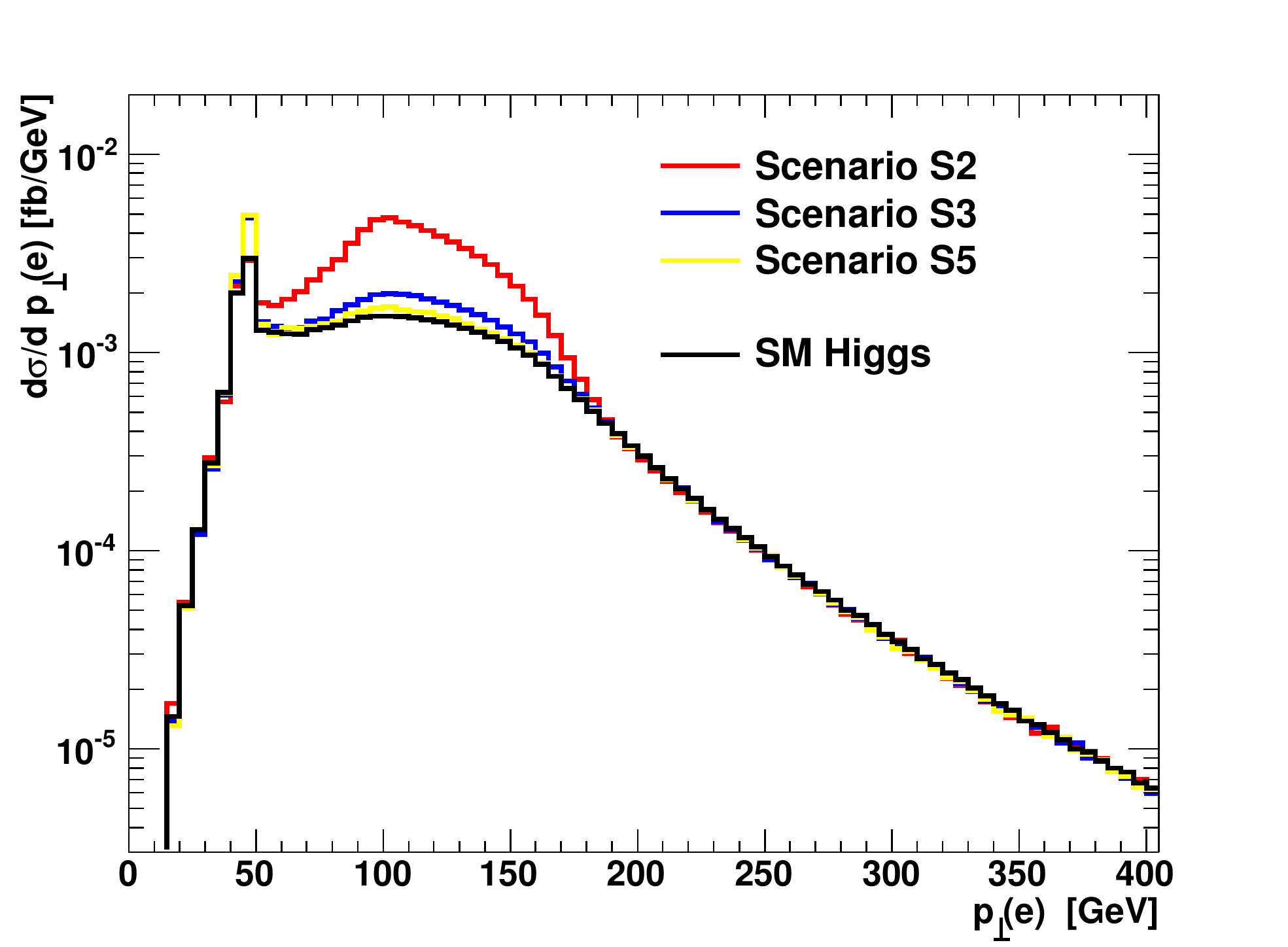} &
\includegraphics[width=0.5\textwidth]{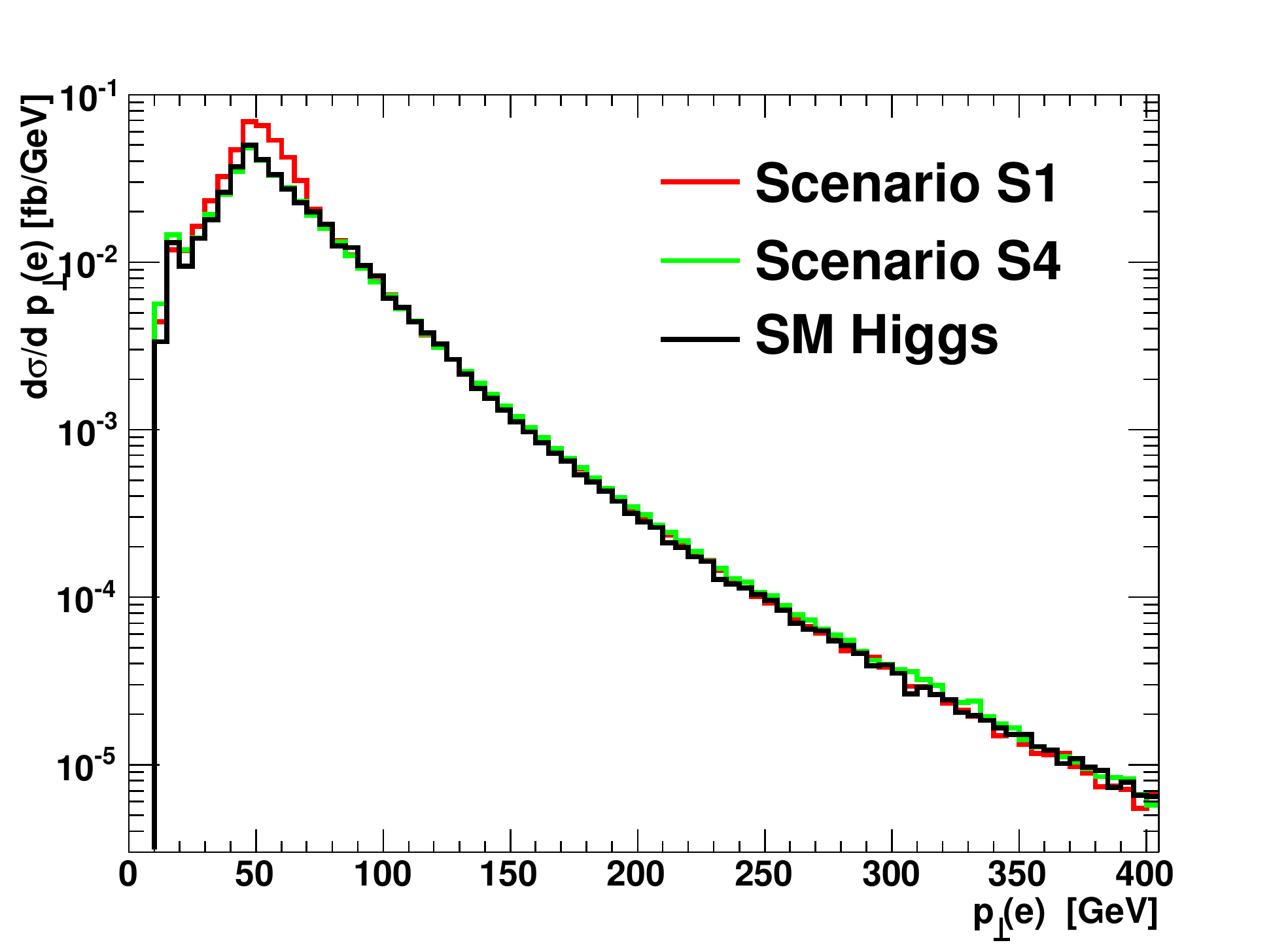} \\[-0.2cm]
 (a) & (b)  \\[-0.1cm]
\includegraphics[width=0.5\textwidth]{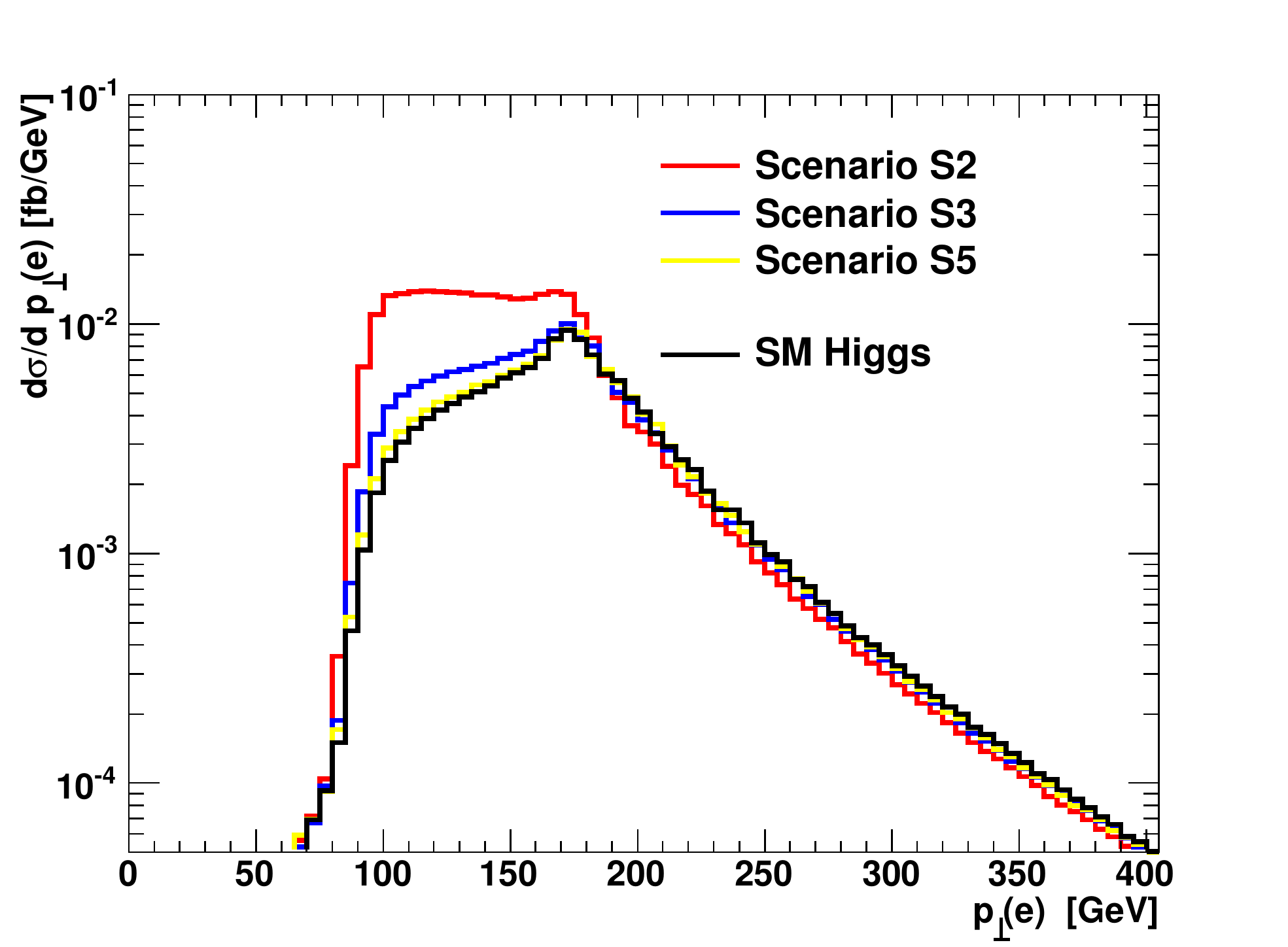} &
\includegraphics[width=0.5\textwidth]{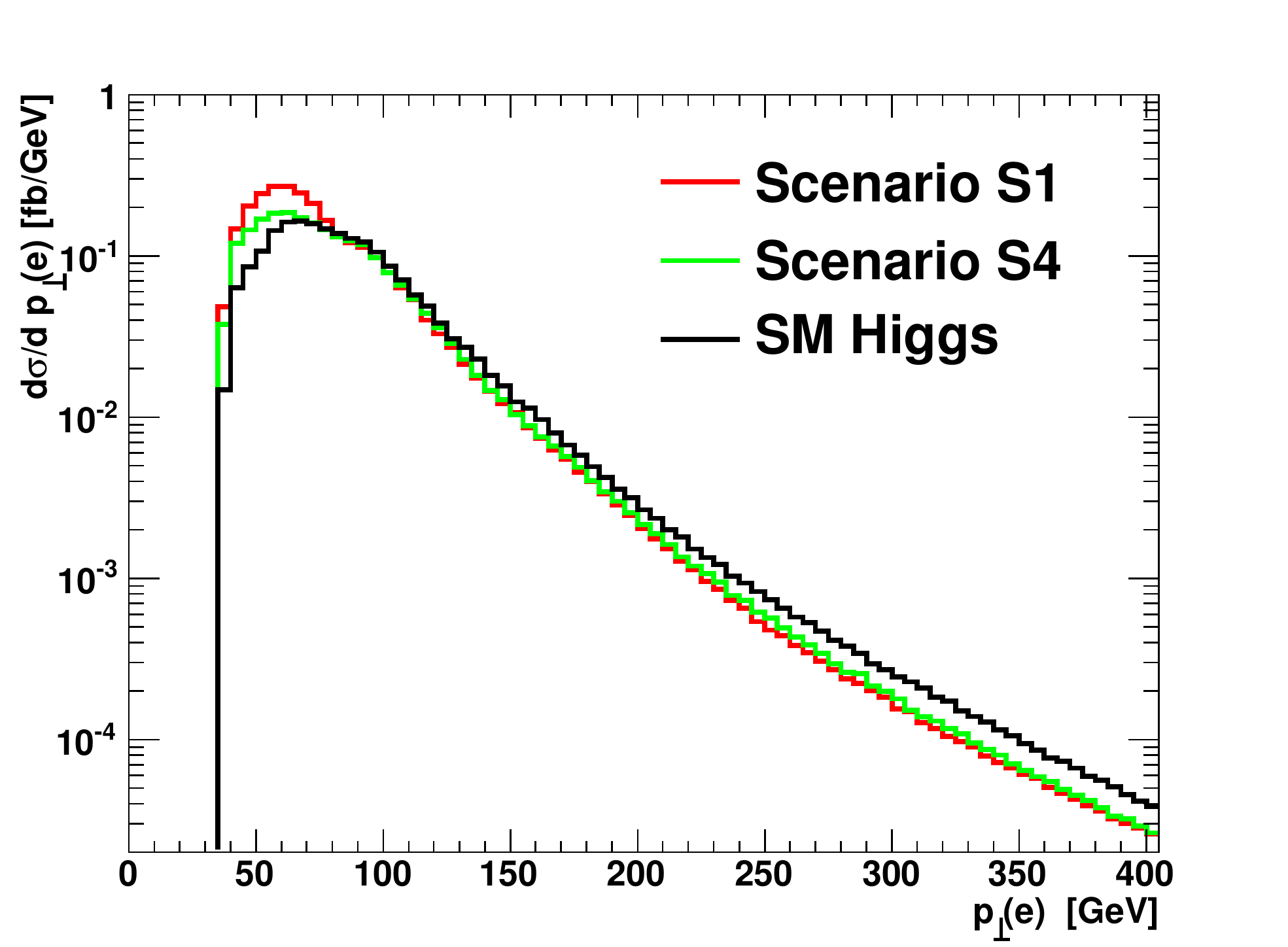}\\[-0.2cm]
 (c) & (d)
\end{tabular}
\end{center}
\vspace{-0.6cm}
\caption{Transverse momentum distribution of the hardest electron (electron or positron) for (a,b) $gg\rightarrow e^{+}e^{-}\mu^{+}\mu^{-}$ 
  and for (c,d) $gg\rightarrow e^{+}e^{-}\nu_l\bar\nu_l$ at $\sqrt{s}=13$\,TeV.
  Due to the different cuts on the invariant mass of the four leptons, the scenarios S2, S3
  and S5 (a,c) are plotted separately from S1 and S4 (b,d).}
\label{fig:pte} 
\end{figure}
\setlength{\tabcolsep}{6pt}

In the following paragraphs we also discuss how other observables are affected by the presence of a second heavy Higgs boson.
In the invariant mass distribution the effect is obvious as the heavy Higgs boson leads to an additional peak in the distribution.
In contrast in other observables its presence is less significant but still sizeable.

In \fig{fig:pte} we plot the transverse momentum distribution of the hardest 
electron, being either the involved electron or the positron. In \fig{fig:pte}~(a,b) the distribution is plotted for the final state 
including muons, \fig{fig:pte}~(c,d) shows the same distribution for the neutrino final state. The scenarios S2, S3 and S5 are plotted
in a different figure than the scenarios S1 and S4 due to the different cuts on the invariant mass of the four leptons
($350$\,GeV and $100$\,GeV).
In order to investigate the question how the different \thdm{} benchmark scenarios can be distinguished from
the \sm, we also plot the distribution for the \sm{} including the \sm{} Higgs boson.
In contrast to the invariant mass and transverse mass spectra
we do not split our results into the pure heavy Higgs boson and interference contributions, since
the individual contributions would include large invariant mass contributions, not being unitarized.

For the \sm{} with a cut on the invariant mass of $350$\,GeV we observe a peak coming from the $Z$-boson decay 
followed by a smoothly falling distribution, see \fig{fig:pte}~(a,c). Reducing the cut to $100$\,GeV changes the shape of the distribution 
significantly. In case of the scenarios S2, S3 and S5 we observe a substantial deviation from the Standard Model prediction
for the intermediate region $~ 50 - 200$\,GeV, which is caused by the effects of the additional heavy Higgs boson. For the cases of the 
scenarios S1 and S4 the effects are much less pronounced. This is partially due to the decreased importance of the heavy 
Higgs boson but also due to the fact that the heavy Higgs is $200$\,GeV in these scenarios, which means that the cut on the invariant
mass is set to $100$\,GeV. These distributions therefore contain contributions from the light Higgs peak,
which makes the heavy Higgs also relatively unimportant.
For the neutrino final state one does not observe a peak in the transverse momentum distribution in the low $p_T$-region,
which can be explained by the additional cut on the missing $E_T$ of the neutrinos and by the additional presence of 
the $W$-pair processes plus the increased importance of off-shell contributions.

\setlength{\tabcolsep}{0pt}
\begin{figure}[h!]
\begin{center}
\begin{tabular}{cc}
\includegraphics[width=0.45\textwidth]{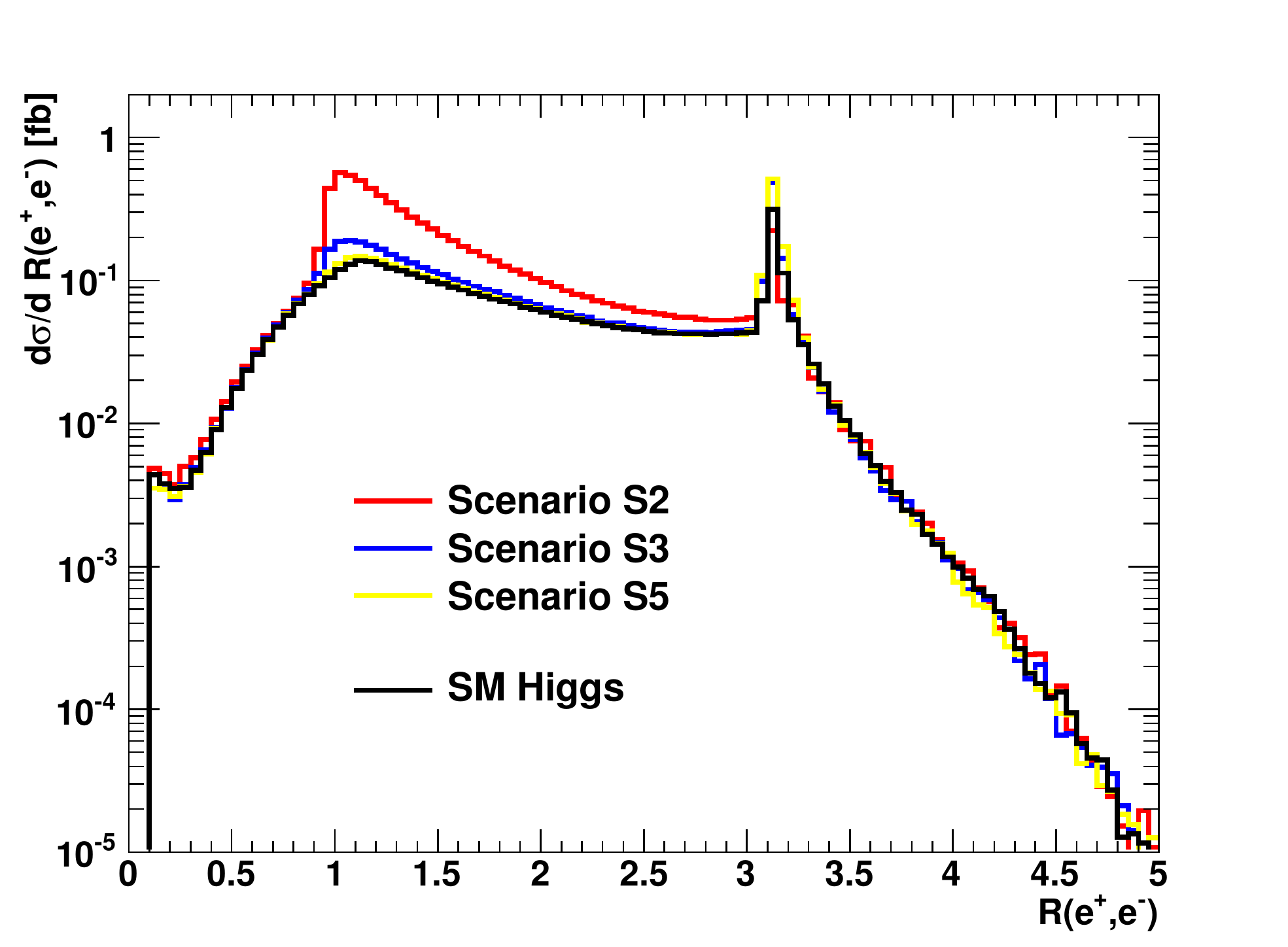} &
\includegraphics[width=0.45\textwidth]{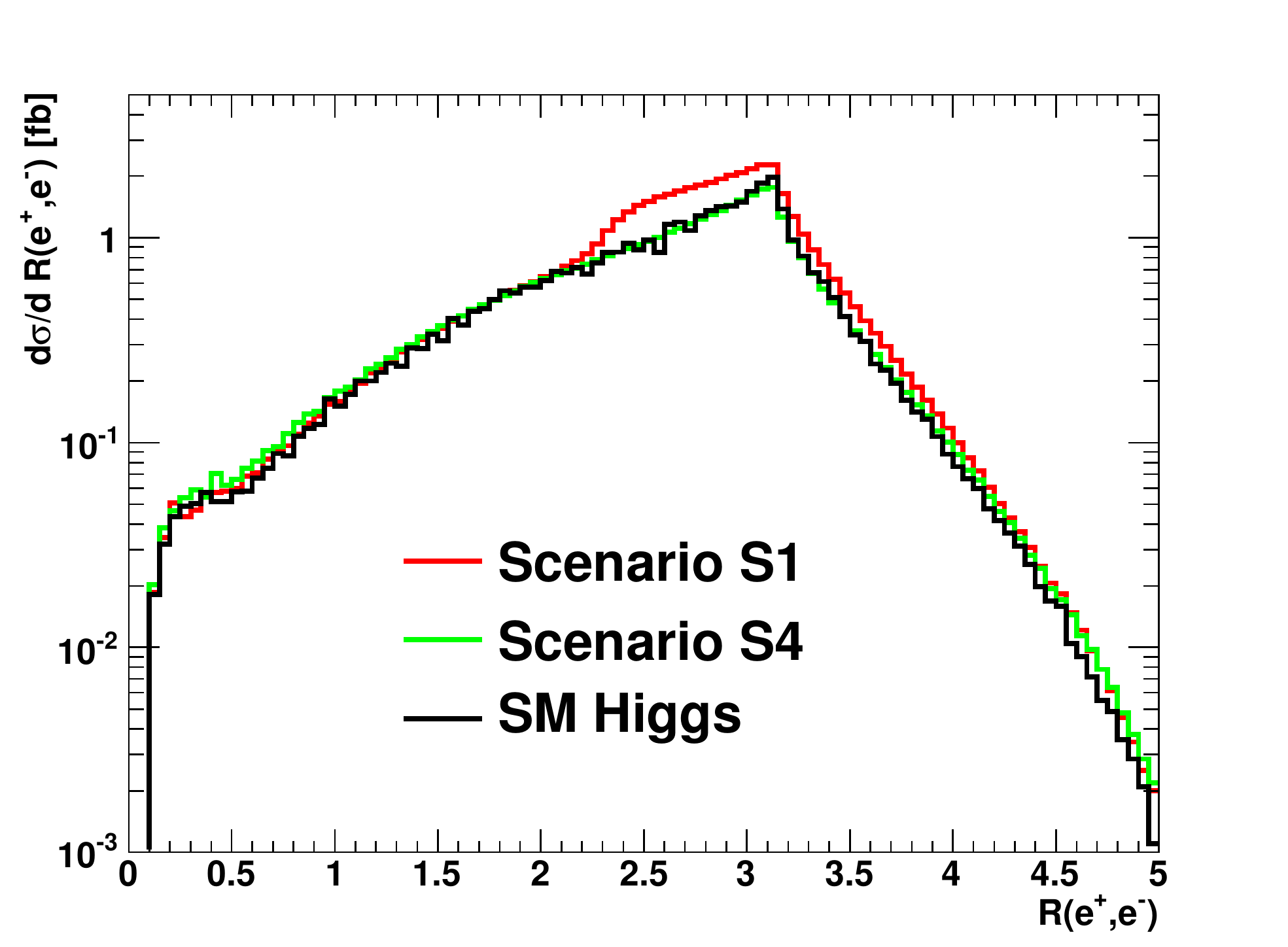}\\[-0.2cm]
 (a) & (b)  \\[-0.1cm]
\includegraphics[width=0.45\textwidth]{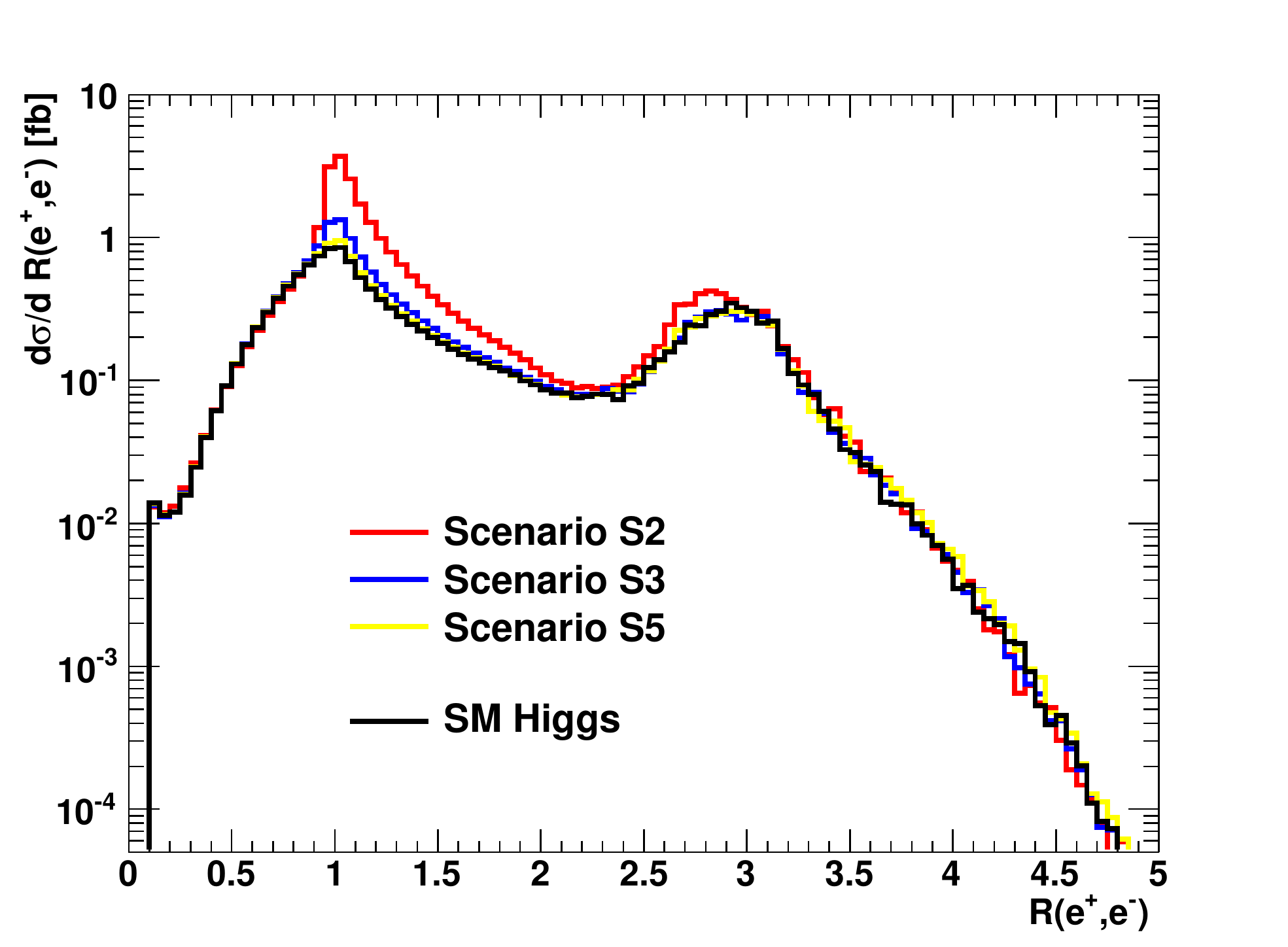} &
\includegraphics[width=0.45\textwidth]{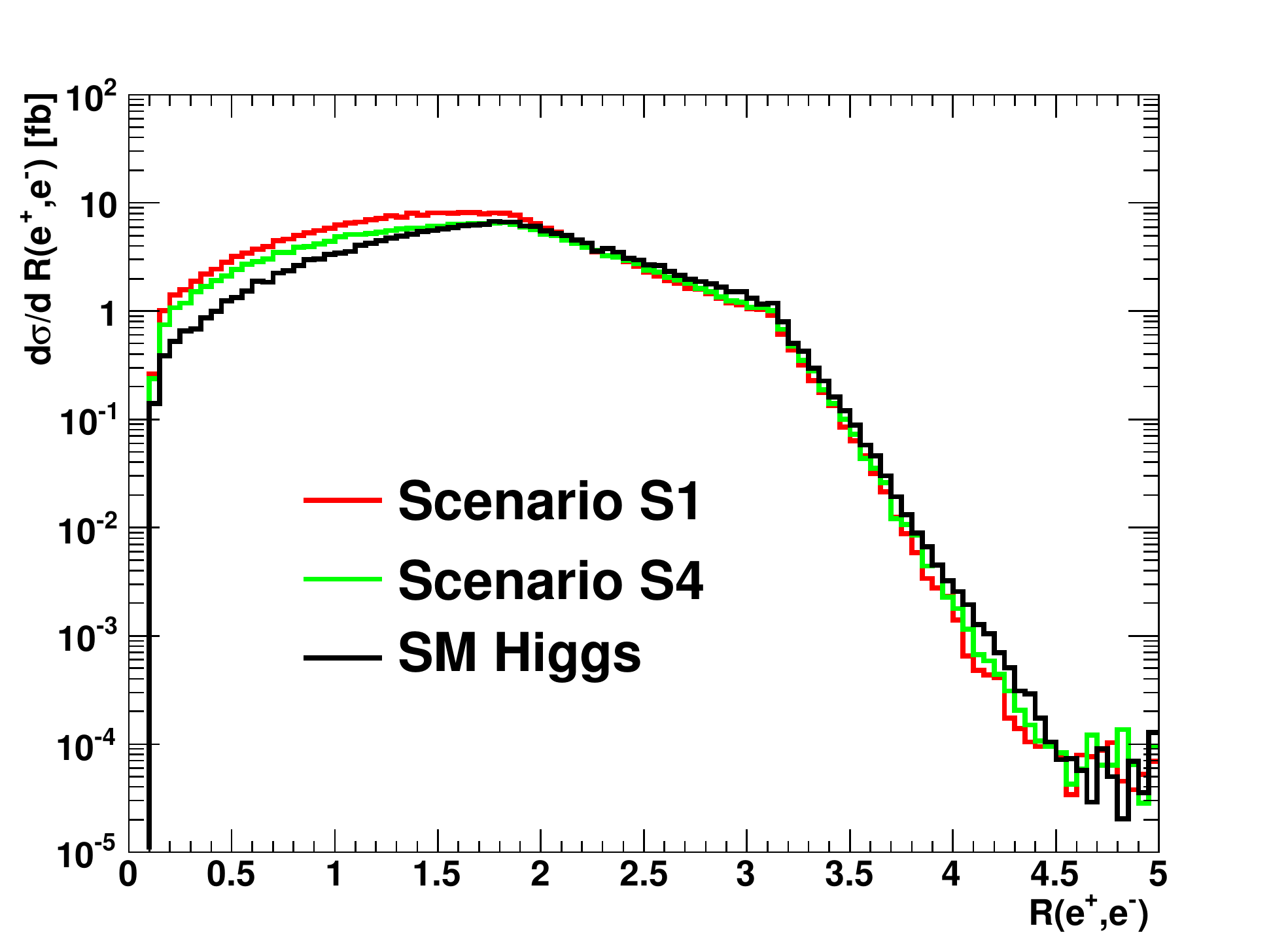}\\[-0.2cm]
 (c) & (d)
\end{tabular}
\end{center}
\vspace{-0.6cm}
\caption{$R$-separation between electron and positron for (a,b) $gg\rightarrow e^{+}e^{-}\mu^{+}\mu^{-}$ 
  and for (c,d) $gg\rightarrow e^{+}e^{-}\nu_l\bar\nu_l$ at $\sqrt{s}=13$\,TeV.
  Due to the different cuts on the invariant mass of the four leptons, the scenarios S2, S3
  and S5 (a,c) are plotted separately from S1 and S4 (b,d).}
\label{fig:Ree} 
\end{figure}
\setlength{\tabcolsep}{6pt}

Another interesting class of observables are angular correlations. In \fig{fig:Ree} we show the $R$-separation between the 
electron and the positron. Also here we have split the plots according to the different cuts in the same way as described above.

For $gg\rightarrow e^{+}e^{-}\mu^{+}\mu^{-}$ we observe a peak of the distribution at $\pi$ which stems from $Z$-boson decays at rest
where the leptons are in a back-to-back configuration. Adding an additional heavy Higgs boson leads to $Z$ bosons that are boosted, which
means that in the lab frame the distance between the same flavor leptons is reduced. The presence of a heavy Higgs therefore 
tends to shift the $R$-separation from back-to-back configuration toward smaller values. For the scenarios~S1 and S4 this
effect is much less pronounced as the importance of the heavy Higgs is reduced and in addition there is a contribution from the 
light Higgs (due to the lower cut on the invariant mass), which, however, cannot lead to boosted Z bosons. 
Therefore we do not observe this shift toward
smaller values of $R$ but the distribution peaks at $R=\pi$.

For $gg\rightarrow e^{+}e^{-}\nu_l\bar\nu_l$ the peak at $\pi$ is reduced to a kink, again due to the presence of an additional $W$-pair channel,
but also here we see the tendency that the presence of the heavy Higgs shifts the $R$-separation toward smaller values.

We note that the presence of a heavy Higgs does not lead to drastic shape distortions of angular correlations and observables
of the final state leptons. As both Higgs bosons are scalars, they decay isotropically which means that the heavy Higgs boson influences
rather the kinematics of the leptons than the qualitative decay structure.

%\newpage
\subsection{Relevance of interference contributions exemplified with $gg\rightarrow ZZ$}
\label{sec:ggZZ}

In the following study of the relevance of interference contributions we make use
of the five benchmark scenarios presented in \tab{tab:2hdm}
and vary one of the three relevant parameters for our process, namely $\mH$, $\sin(\beta-\alpha)$ and $\tb$.
The combination of the latter two also fixes the coupling of
the heavy Higgs to top quarks and bottom quarks, see \eqn{eq:gtbH}.

\subsubsection{Dependence on $\mH$ in scenarios S1}
\label{sec:S1mH}

\setlength{\tabcolsep}{0pt}
\begin{figure}[!h]
\begin{center}
\begin{tabular}{cc}
\includegraphics[width=0.47\textwidth]{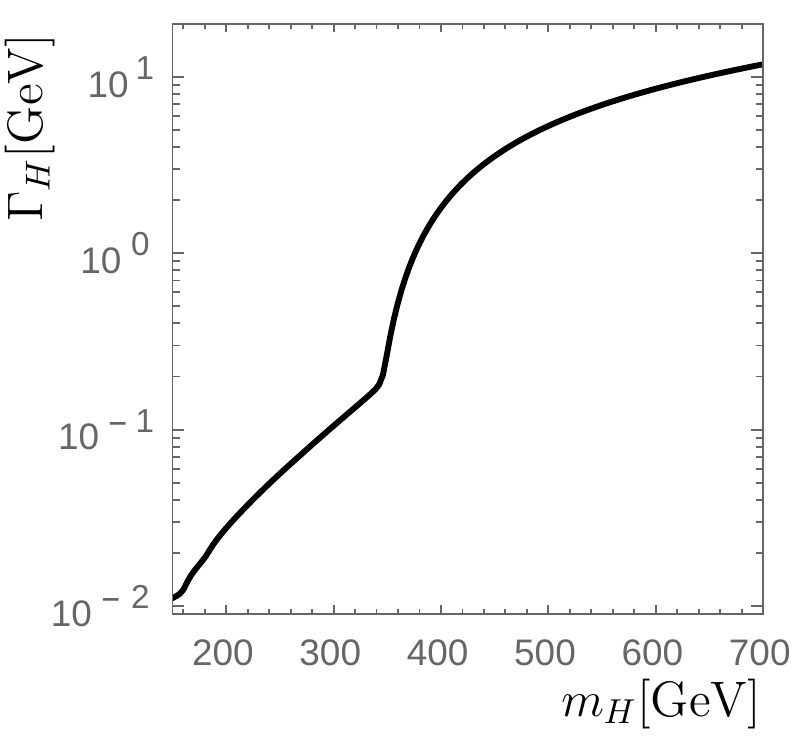} & 
\includegraphics[width=0.47\textwidth]{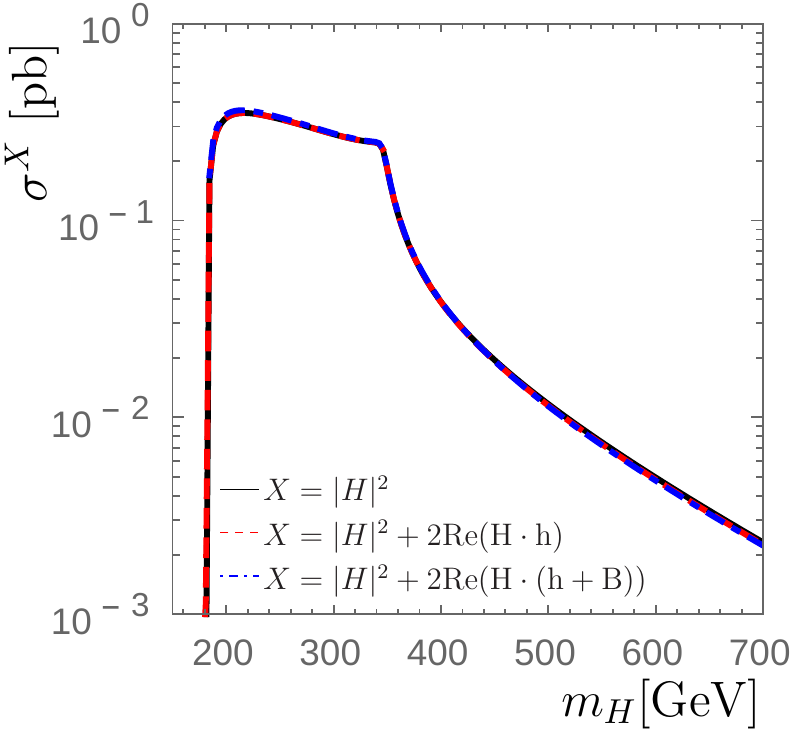} \\[-0.5cm]
 (a) & (b) \\
\multicolumn{2}{c}{\includegraphics[width=0.47\textwidth]{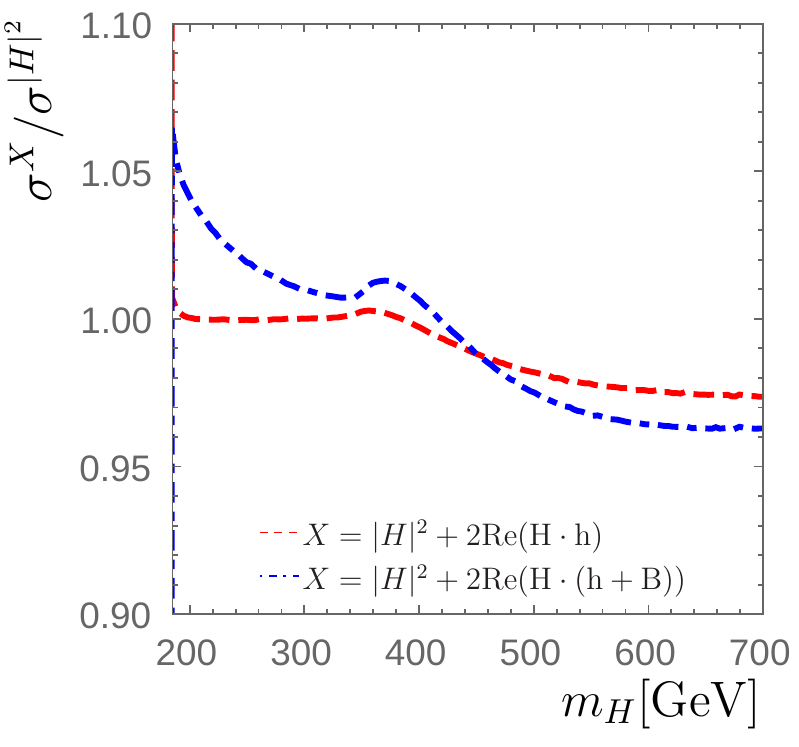}} \\[-0.7cm]
\multicolumn{2}{c}{(c)}
\end{tabular}
\end{center}
\vspace{-0.6cm}
\caption{Scenario S1 as a function of the heavy Higgs mass $\mH$ in GeV showing (a) the Higgs width $\GaH$ in GeV;
(b) the inclusive cross section $\sigma^X$ in pb within $\mzz^I$ and for $\sqrt{s}=8$\,TeV 
(black: $X=|H|^2$; red, dashed: $X=|H|^2+2\text{Re}(H\cdot h)$;
blue, dot-dashed: $X=|H|^2+2\text{Re}(H\cdot h)+2\text{Re}(H\cdot B)$);
(c) the relative ratio of cross sections $\sigma^{X}/\sigma^{|H|^2}$
within $\mzz^I$.}
\label{fig:S1mH} 
\end{figure}
\setlength{\tabcolsep}{6pt}

In order to make contact with the ATLAS analysis carried out in \citere{Aad:2015kna}
we start with a discussion of the interference contributions as a function of $\mH$ in scenario~S1.
In case of S1 the variation within $\mH=150-700$\,GeV corresponds to Fig. 14 
of \citere{Aad:2015kna},
where we keep $\cba=-0.1$ (in the convention for $\beta-\alpha$ of the ATLAS analysis)
fixed (as in Fig. 14 of \citere{Aad:2015kna})
and pick one value of $\tb$, namely $\tb=2$, which leads to cross sections
close to the achieved experimental sensitivity in the first run of the \lhc{}.
To match the numbers of the ATLAS analysis the centre-of-mass energy for scenario S1 is set to $\sqrt{s}=8$\,TeV.
We quantify the relevance of the interference as follows: We integrate the differential cross section
$d\sigma^X/d\mzz$ for $gg\rightarrow ZZ$ in the range $\mzz\in \mzz^I=[\mH-15\,\text{GeV},\mH+15\,\text{GeV}]$ 
and compare the pure heavy Higgs signal peak $X=|H|^2$ with the heavy Higgs peak taking into
account the interference with the light Higgs $X=|H|^2+2\text{Re}(H\cdot h)$ and the
interference with the light Higgs and the background $X=|H|^2+2\text{Re}(H\cdot h)+2\text{Re}(H\cdot B)$,
where the background stems from the box diagrams not involving a Higgs boson.
All individual contributions are gauge-invariant in this case.
The procedure does not take into account the fact that the interferences are often
close to be point symmetric with respect to $\mzz=\mH$, i.e.\ parts of the interferences below
($\mzz<\mH$) and above ($\mzz>\mH$) the heavy Higgs boson peak cancel each other
when integrating over the whole range $\mzz^I$.
This is however only of relevance if the peak can be experimentally resolved and
of large importance in the case the mass (and possibly width) of a heavy Higgs should be deduced
from the peak position and its structure. For a discussion of axially and point symmetric contributions
with respect to $\mzz=\mH$ we also refer to \citere{Jung:2015sna}, where the imaginary
part of the contributions, due to its axially symmetric structure, is identified to be most relevant
for interferences.

\fig{fig:S1mH}~(a) shows the Higgs width~$\GaH$
as a function of the Higgs mass $\mH$ for scenario~S1. The width~$\GaH$ increases rapidly
above the top-quark threshold $\mH>2m_t$,
which lowers the inclusive cross section within $\mzz^I$ significantly, as it can be deduced
from \fig{fig:S1mH}~(b). The latter presents $\sigma^X$ for the pure signal $X=|H|^2$ in black as well as
for $X=|H|^2+2\text{Re}(H\cdot h)$ in red and for $X=|H|^2+2\text{Re}(H\cdot h)+2\text{Re}(H\cdot B)$ in blue.
The relative difference of the contributions with respect to $|H|^2$ is shown in \fig{fig:S1mH}~(c).
We deduce that the interference contributions are below $\mathcal{O}(10\%)$ in the whole
parameter range, even in regions where the Higgs width becomes so large that the validity of 
the narrow-width approximation (NWA) for $H$ can be questioned.
Note that the interferences are getting larger for $gg\rightarrow ZZ$, when the Higgs mass approaches the
kinematic threshold $2\mz$ from above, since in this example the interferences are
point symmetric with respect to $\mzz=\mH$ and the phase space region $2\mz<\mzz<\mH$ is reduced.
Still we conclude that the usage of the NWA for $H$ 
in the ATLAS analysis \cite{Aad:2015kna} is justified
given the obtained sensitivities on the cross sections.
Similar conclusions can also be drawn in the mentioned ``flipped Yukawa'' scenario, even though
it allows for slightly larger values of $\cba$.

Given the presented results the width and mass of the heavy Higgs are not the most relevant parameters for what concerns
the size of the interferences.
As we will see in the two subsequent subsections an increase in the relevance
of the interference contributions is observed if the couplings of the heavy Higgs boson, in particular
to the top-quark and bottom-quark loop, are suppressed or the latter is significantly enhanced.

\subsubsection{Dependence on $\sba$ in scenarios S1 and S4}

In this subsection we treat $\sba (\cba)$ as the parameter that is
varied and set $\tb$ to fixed values. For what concerns scenario~S1, this
choice corresponds to a line in Fig. 13 of \citere{Aad:2015kna} if we fix $\tb=2$.
Second we pick the type I scenario~S4.
In both cases we vary $\cba^2$ in the range $0.00001-1$, which corresponds
to $\sba=0-0.999995$.
We quantify the interferences as done in the previous subsection again for
$\sqrt{s}=8$\,TeV for S1 and afterwards for $\sqrt{s}=13$\,TeV for S4.

\setlength{\tabcolsep}{0pt}
\begin{figure}[htp]
\begin{center}
\begin{tabular}{cc}
\includegraphics[width=0.47\textwidth]{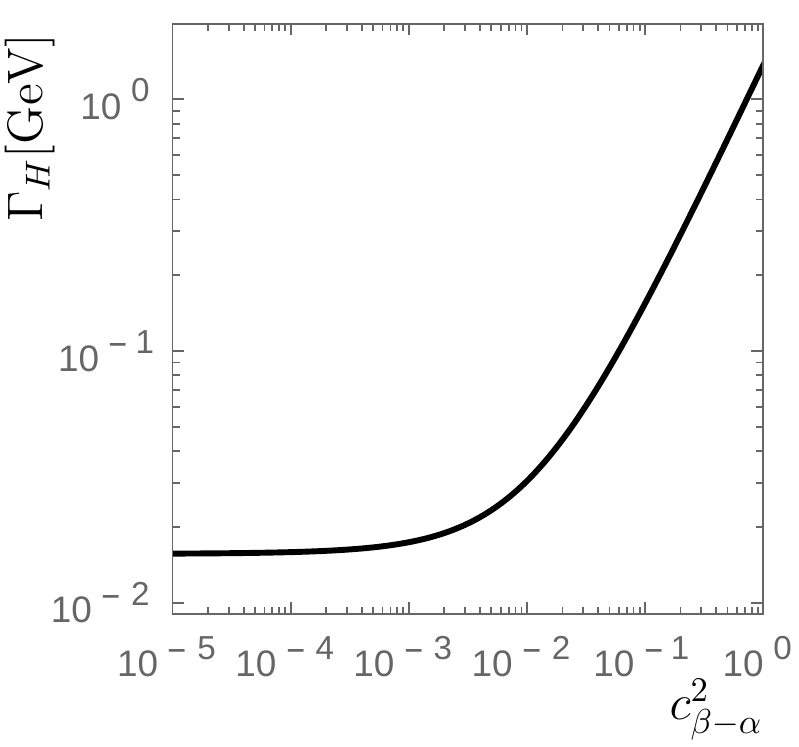} & 
\includegraphics[width=0.47\textwidth]{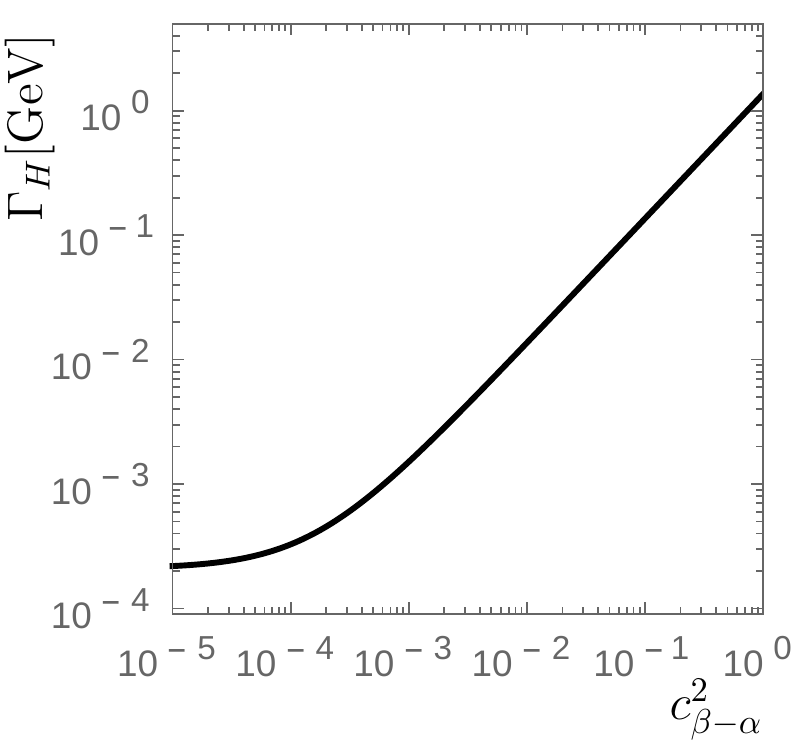} \\[-0.5cm]
 (a) & (b) \\
\includegraphics[width=0.47\textwidth]{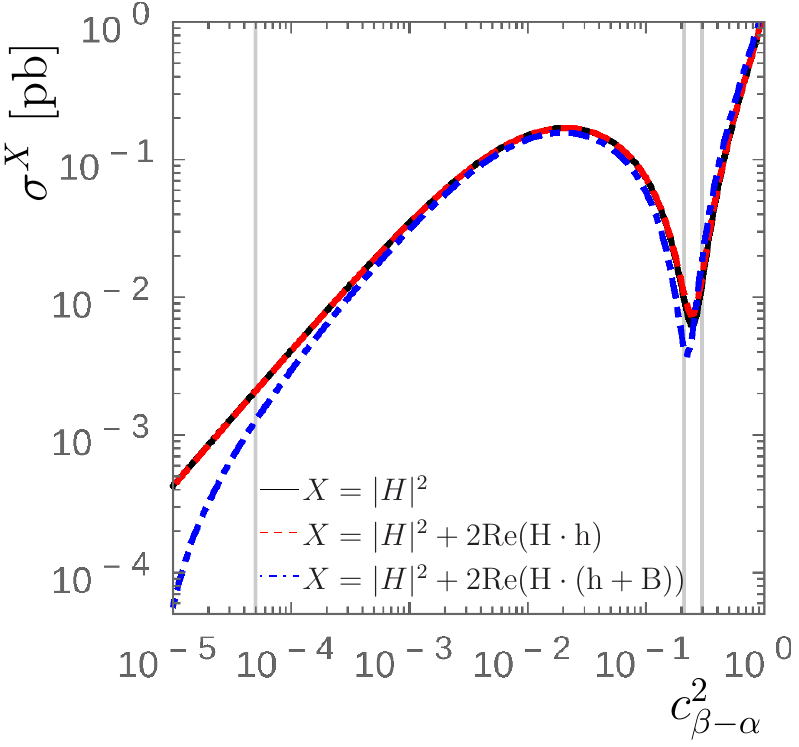} & 
\includegraphics[width=0.47\textwidth]{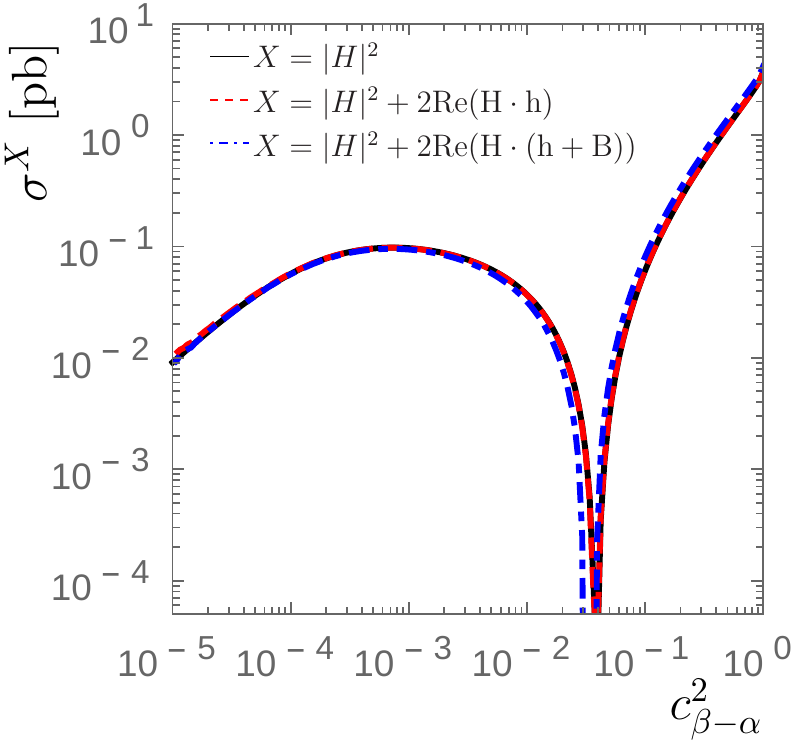} \\[-0.5cm]
 (c) & (d) \\
\includegraphics[width=0.47\textwidth]{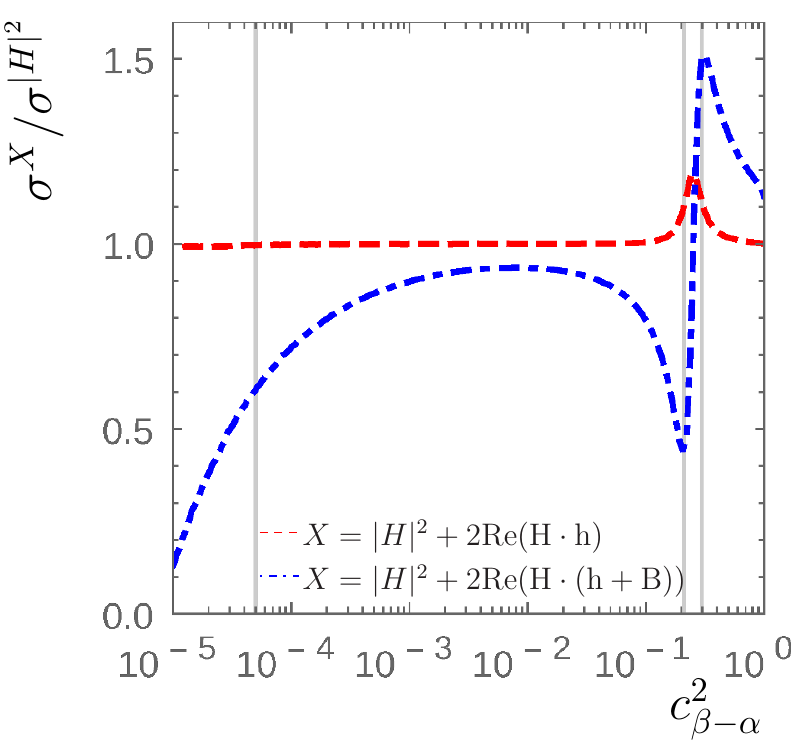} &
\includegraphics[width=0.47\textwidth]{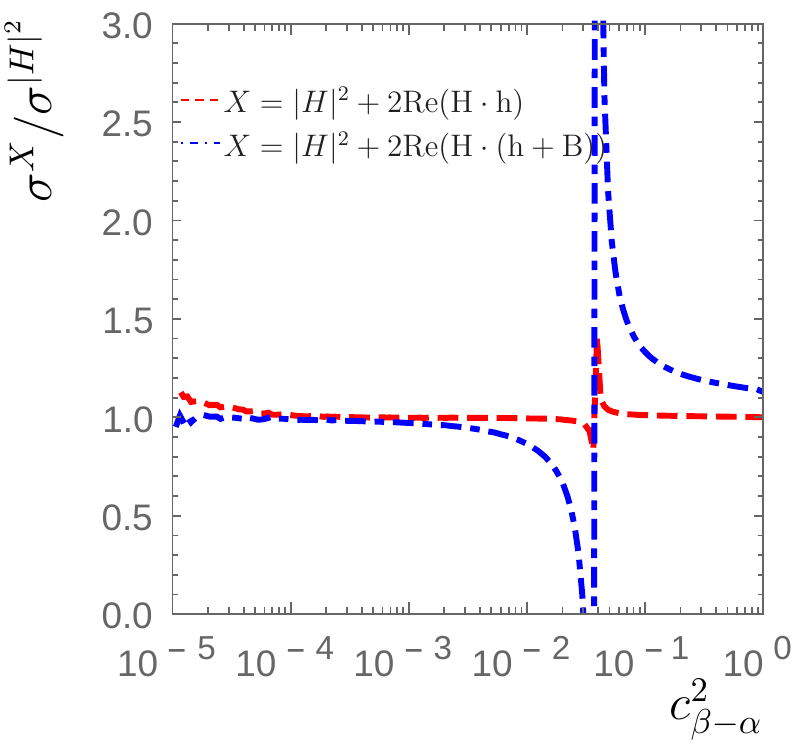} \\[-0.5cm]
 (e) & (f) 
\end{tabular}
\end{center}
\vspace{-0.6cm}
\caption{Scenario S1 (a,c,e) and scenario S4 (b,d,f) as a function of $\cba^2$ showing (a,b) the Higgs width $\GaH$ in GeV;
(c,d) the inclusive cross section $\sigma^X$ in pb within $\mzz^I$ for $\sqrt{s}=8$\,TeV
and $\sqrt{s}=13$\,TeV respectively (black: $X=|H|^2$; red, dashed: $X=|H|^2+2\text{Re}(H\cdot h)$;
blue, dot-dashed: $X=|H|^2+2\text{Re}(H\cdot h)+2\text{Re}(H\cdot B)$); (e,f) the relative ratio of cross sections $\sigma^{X}/\sigma^{|H|^2}$
within $\mzz^I$. We show $d\sigma^X/d\mzz$ at the three marked values of $\cba^2$ in (c,e) in \fig{fig:S1diff}.}
\label{fig:S1S4cba} 
\end{figure}
\setlength{\tabcolsep}{6pt}

\begin{figure}[htp]
\begin{center}
\begin{tabular}{ccc}
\includegraphics[width=0.33\textwidth]{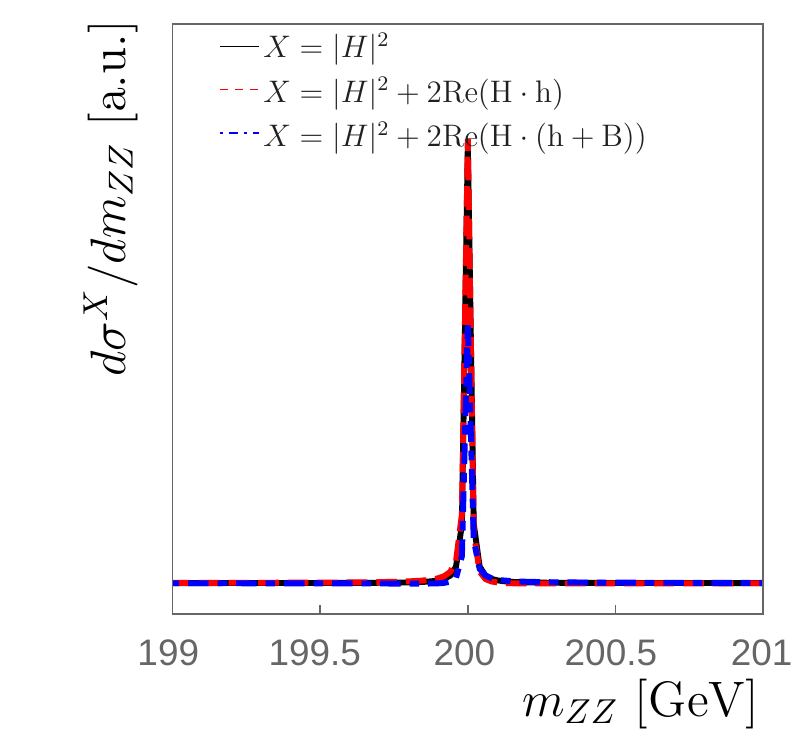} & 
\includegraphics[width=0.33\textwidth]{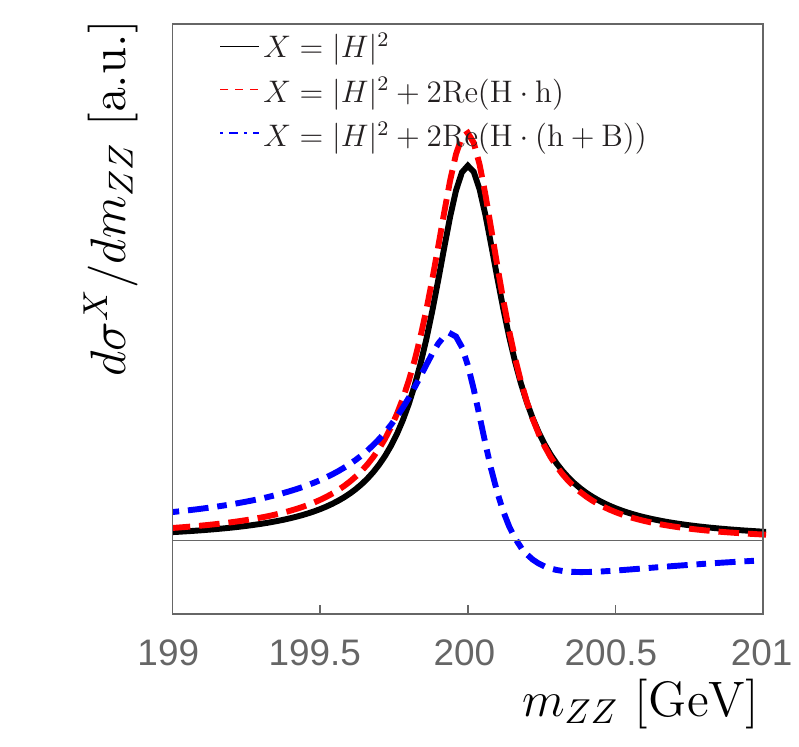} & 
\includegraphics[width=0.33\textwidth]{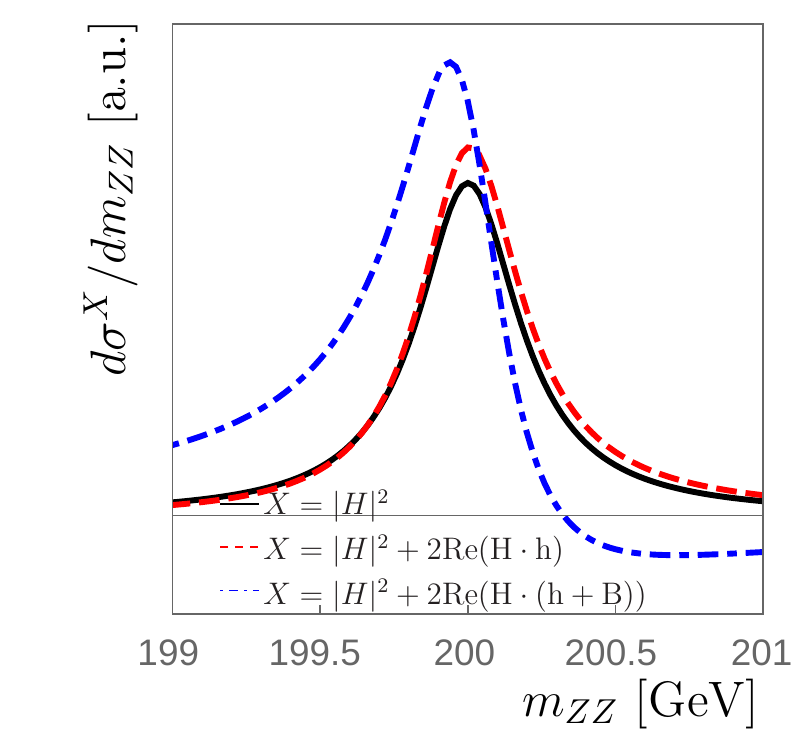} \\[-0.5cm]
 (a) & (b) & (c)
\end{tabular}
\end{center}
\vspace{-0.6cm}
\caption{Partonic cross sections $d\sigma^X/d\mzz$ in arbitrary units (see text) as a function of $\mzz$ in GeV
for S1 for three different values of $\cba^2$ marked in \fig{fig:S1S4cba}~(c,e), namely
(a) $\cba^2=0.00005$; (b) $\cba^2=0.21$ and (c) $\cba^2=0.2944$. The color coding again corresponds to
black: $X=|H|^2$; red, dashed: $X=|H|^2+2\text{Re}(H\cdot h)$;
blue, dot-dashed: $X=|H|^2+2\text{Re}(H\cdot h)+2\text{Re}(H\cdot B)$.}
\label{fig:S1diff} 
\end{figure}

In \fig{fig:S1S4cba}~(a,c,e) we present our results for S1. The figures provide complementary
information as \fig{fig:S1mH}.
Two interesting observations can be made: The interference is naturally larger in regions where
the signal cross section $|H|^2$ becomes small as can be seen for very low values of $\cba^2$.
In addition, and this effect is generic to the \thdm{}, the coupling of the heavy Higgs to individual quarks 
can vanish or the contributions between the top- and bottom-quark cancel.
In our example at $\cba^2=0.2$ the coupling to top quarks tends to zero, which can
easily be understood by decomposing $g_t^H=\sin\alpha/\sin\beta=-\sba/\tb+\cba$.
For $\tb>0.7$ the corresponding value of $\sba$, where $g_t^H=0$, is above $\sba>0.57$ approaching
$1$ with increasing $\tb$.
Necessarily the interferences in those regions appear to be large.
Due to the remaining bottom-quark contributions the minimum is actually located
at $\cba^2=0.23$ in this specific example.
The occurrence of large interferences however appears in regions where the signal cross sections
is below experimental sensitivities, which reached $\mathcal{O}(0.05-1\,\text{pb})$ 
for the process $gg\rightarrow ZZ$ according to Fig. 12 in \citere{Aad:2015kna}
in the first run of the \lhc{}.
The observed large effects of interferences occur at lower cross sections,
which are however potentially of relevance with increasing statistics at the \lhc{}.
We show the different distributions $d\sigma^X/d\mzz$ at three values of $\cba$ (marked in \fig{fig:S1S4cba}~(c,e))
in \fig{fig:S1diff}. Since we obtained the results of \fig{fig:S1diff} for the partonic cross section
and are interested in the relative contributions of the interference effects with respect to the
pure heavy Higgs boson signal,
we omit units at the $y$-axis. The inclusive cross section can be deduced from \fig{fig:S1S4cba}~(c).
The three values of $\cba^2$ correspond to $\sba=0.999975,0.8888$ and $0.84$.

In \citere{Liebler:2015aka} the relevance of interferences for a heavy Higgs boson in $e^+e^-\rightarrow ZVV/\nu\bar\nu VV$
in the context of a \thdm{} at a linear collider was discussed. Since in these processes the Higgs is
produced and decays through the couplings to heavy gauge bosons, the coupling $g_V^H$ occurs twice in
each Feynman diagram, whereas $gg\rightarrow ZZ$ only comes with one appearance.
However, vector-boson fusion at the \lhc{} also shows a larger suppression of the heavy Higgs boson
signal involving the relevant coupling twice.
The larger suppression naturally induces larger interference contributions, but however
small event rates.

\setlength{\tabcolsep}{0pt}
\setlength{\tabcolsep}{6pt}

We finally discuss in this context a \thdm{} of type~I.
As we have seen already a crucial quantity
is the top-quark Yukawa coupling of the heavy Higgs~$g_t^H$.
Since in a \thdm{} of type~I the relative bottom-quark and top-quark Yukawa
coupling are equal, $g_t^H=g_b^H$,
the cross section as a function of $\cba^2$ does indeed vanish for one specific value of $0.57<\sba\leq 1$
rather than just showing a minimum as depicted in \fig{fig:S1S4cba}~(c). We show this behaviour in
\fig{fig:S1S4cba}~(b,d,f) for scenario~S4, which again gives rise to very large interferences in this region,
however with low cross sections below $10^{-1}$\,pb.

\subsubsection{Dependence on $\tb$ in S2}

\begin{figure}[!h]
\begin{center}
\begin{tabular}{cc}
\includegraphics[width=0.47\textwidth]{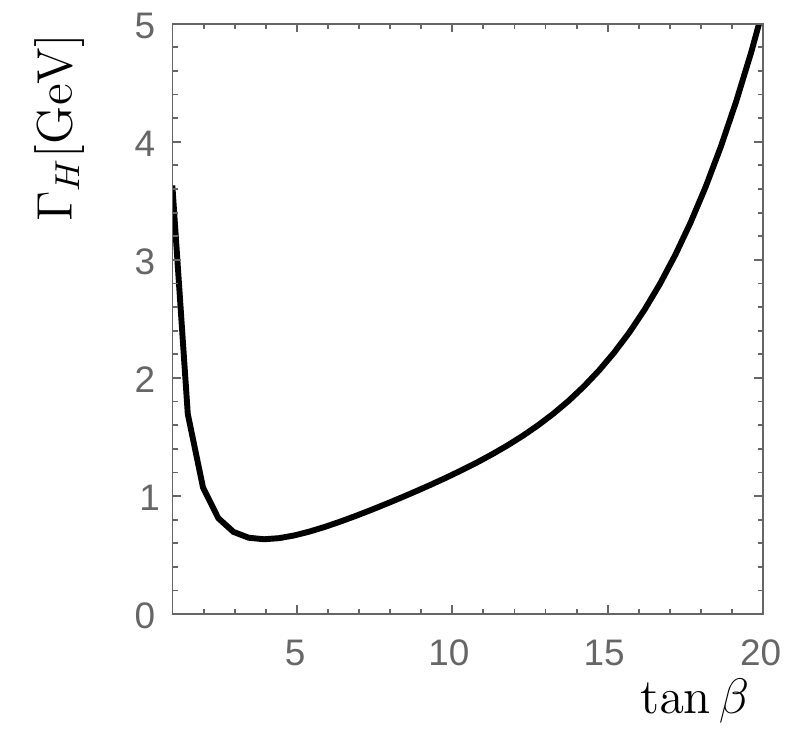} & 
\includegraphics[width=0.47\textwidth]{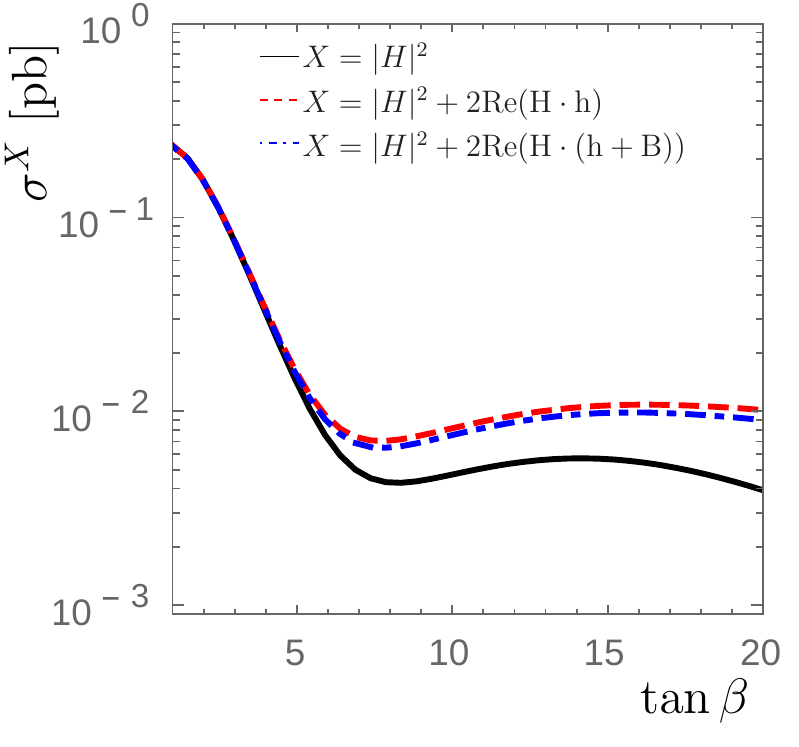}  \\[-0.7cm]
 (a) & (b)  \\
\includegraphics[width=0.47\textwidth]{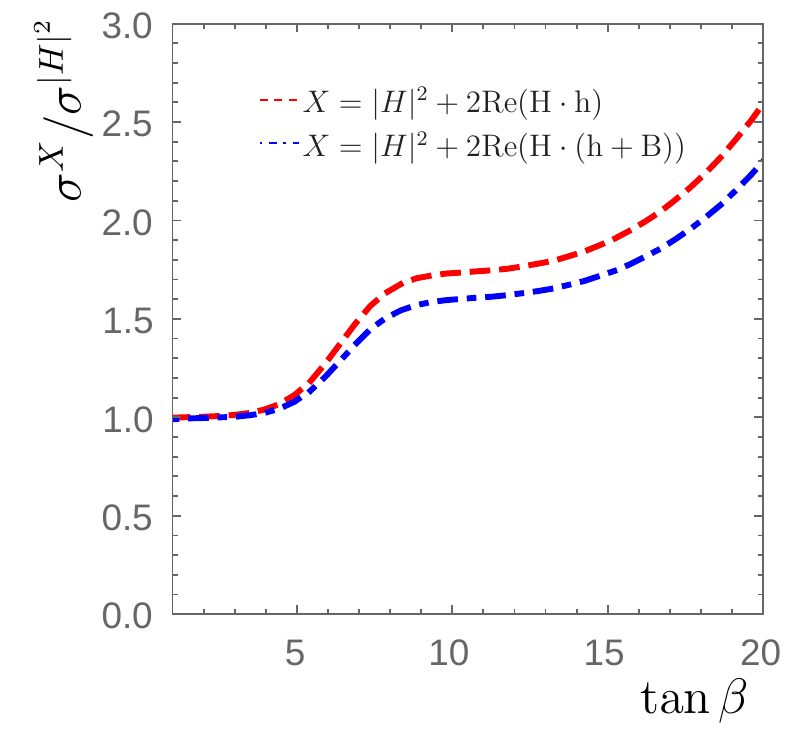} &
\includegraphics[width=0.47\textwidth]{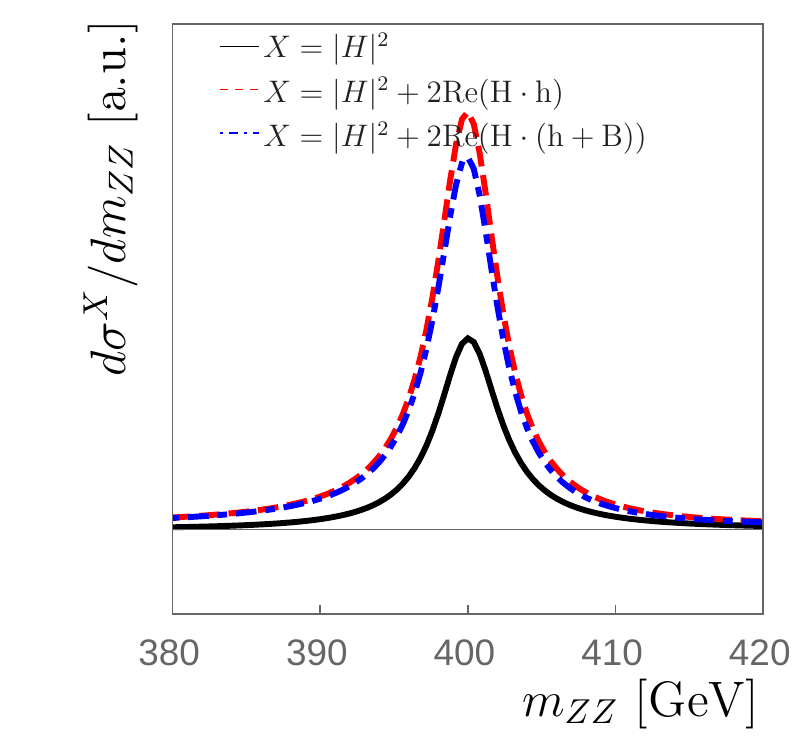}  \\[-0.7cm]
 (c) & (d) 
\end{tabular}
\end{center}
\vspace{-0.6cm}
\caption{Scenario S2 (S5) as a function of $\tb$ showing (a) the Higgs width $\GaH$ in GeV;
(b) the inclusive cross section $\sigma^X$ in pb within $\mzz^{I}$ for $\sqrt{s}=13$\,TeV
(black: $X=|H|^2$; red, dashed: $X=|H|^2+2\text{Re}(H\cdot h)$;
blue, dot-dashed: $X=|H|^2+2\text{Re}(H\cdot h)+2\text{Re}(H\cdot B)$);
(c) the relative ratio of cross sections $\sigma^{X}/\sigma^{|H|^2}$
within $\mzz^{I}$. The partonic cross section $d\sigma^X/d\mzz$ in arbitrary units (see text) is shown
in (d) as a function of $\mzz$ in GeV for scenario S5 (S2 with $\tb=20$).}
\label{fig:S2tb} 
\end{figure}

Lastly we focus on the dependence on $\tb$ in a \thdm{} type II,
where with increasing $\tb$ the coupling to bottom-quarks tends to be enhanced for all Higgs bosons.
\fig{fig:S2tb} shows the Higgs width $\GaH$ and the relevance of the interferences as a function of $\tb$ for scenario~S2.
With increasing $\tb$ the total width~$\GaH$, depicted in \fig{fig:S2tb}~(a), first drops due to the drop
in the partial width $H\rightarrow t\bar t$, for high $\tan\beta$
it rises due to the increase of the partial width~$H\rightarrow b\bar b$.
A fraction of heavy Higgs bosons is also decaying into
a pair of lighter Higgs bosons $H\rightarrow hh$. Since however the latter decay mode
depends on the values of $m_A$ and $\mH^{\pm}$ indirectly through the parameters of the Higgs
potential, we stop at $\tb=20$ where $H\rightarrow hh$ is still sub-dominant and our
results can be considered to a large extent independent of $m_A, m_{H^{\pm}}$ and $m_{12}^2$.
We show the inclusive cross section again within the interval $\mzz^{I}=[\mH-15\,\text{GeV},\mH+15\,\text{GeV}]$
in \fig{fig:S2tb}~(b) and (c) for $\sqrt{s}=13$\,TeV.
The local minimum in \fig{fig:S2tb}~(b) at $\tb\sim 7$ again corresponds to $g_t^H=0$, which also explains the short
rise of the cross section for $\tb>7$.
With increasing $\tb$ the cross section $\sigma^{|H|^2}$ then constantly drops, the interferences
on the other hand quickly gain in size. In addition, we show the form of the interferences
for $\tb=20$ in \fig{fig:S2tb}~(d), which defines scenario~S5.
Again these large interferences occur below the current experimental sensitivity, but 
being above $10^{-2}$\,pb the region is potentially in reach for high statistics at the \lhc{}.
We note that values of $\tb>2$ in combination with $\sba=0.99$ for S2 are 
meanwhile excluded from the non-compatibility with the light Higgs boson signal, see e.g.\ \citere{Haber:2015pua}.
However the qualitative features are the same in phenomenologically viable
scenarios with larger values of $\sba$, as we have numerically checked for $\sba=0.999$.

\setlength{\tabcolsep}{0pt}
\setlength{\tabcolsep}{6pt}

The interference contribution with the light Higgs boson 
in this specific case is always positive both below and above the heavy Higgs mass peak
$\mzz=\mH$, see \fig{fig:S2tb}~(d), and significantly larger than in
the previously discussed scenarios due to
the large bottom-quark contribution to $gg\rightarrow H\rightarrow ZZ$, which interferes
with the bottom- and top-quark contribution to $gg\rightarrow h\rightarrow ZZ$.
In contrast the interference with the background yields a much smaller and
negative contribution to the inclusive cross section within $\mzz^{I}$,
which reflects the unitarization of the cross section for large $\mzz$.
Increasing $\sba$ to $0.999$ yields a similar picture, where the large positive
interference is completely dominated by the interference of the bottom-quark
contribution to $gg\rightarrow H\rightarrow ZZ$ with the top-quark contribution
to $gg\rightarrow h\rightarrow ZZ$, however the total inclusive cross section
is reduced to values slightly below $10^{-2}$\,pb. \citere{Jung:2015sna} 
did not point out the relevance of the interference of the heavy Higgs signal
with the light Higgs signal for large values of $\tan\beta$, however emphasized
the importance of the bottom-quark loop in $gg\rightarrow H\rightarrow ZZ$
for what concerns the interference with the background.
Lastly we comment on the influence of the heavy Higgs boson mass.
Below the threshold of the $H\rightarrow hh$ decays, i.e. $\mH<250$\,GeV,
the size of interference with
the light Higgs in $gg\rightarrow ZZ$ is also diminished due to the increase
of $\sigma^{|H|^2}$. The negative interference with the background gets sizeable 
and reduces the cross section by about $50\%$.
Above $\mH>2m_h$, however, $\sigma^X/\sigma^{|H|^2}$ always significantly differs from $1$.
The interferences in these regions can thus significantly enhance the sensitivity
to the heavy Higgs boson in experimental searches.

In total we conclude that in particular for large values of $\tb$ or vanishing $g_t^H$
interferences can get of importance for future experimental analyses. In the first case the interference
of the heavy Higgs contribution with the light Higgs can be significantly enhanced,
in the second case the interference with the background.
Those cases appear in regions where the inclusive cross sections are in the vicinity
of $10^{-2}$\,pb and thus potentially in reach with higher statistics at the \lhc{}.

\subsubsection{Interferences at high invariant masses}
\begin{figure}[htb]
\begin{center}
\begin{tabular}{cc}
\includegraphics[width=0.47\textwidth]{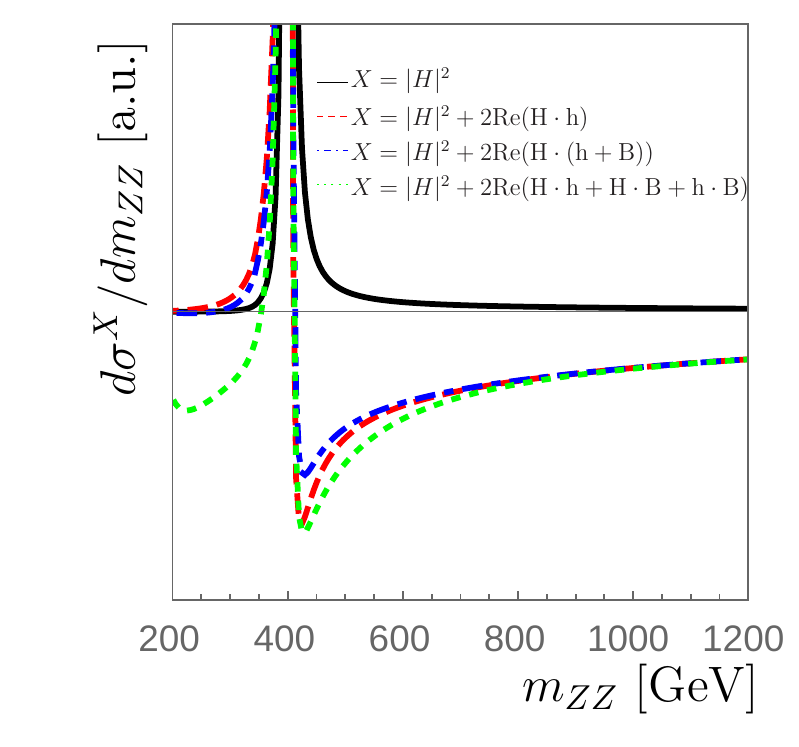} & 
\includegraphics[width=0.47\textwidth]{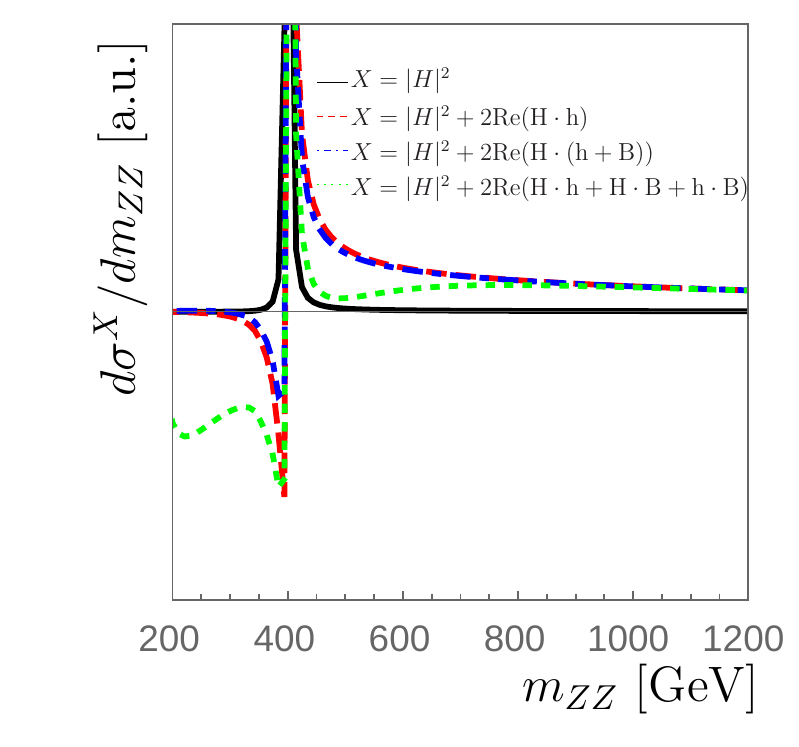} \\[-0.7cm]
 (a) & (b) \\
\multicolumn{2}{c}{\includegraphics[width=0.47\textwidth]{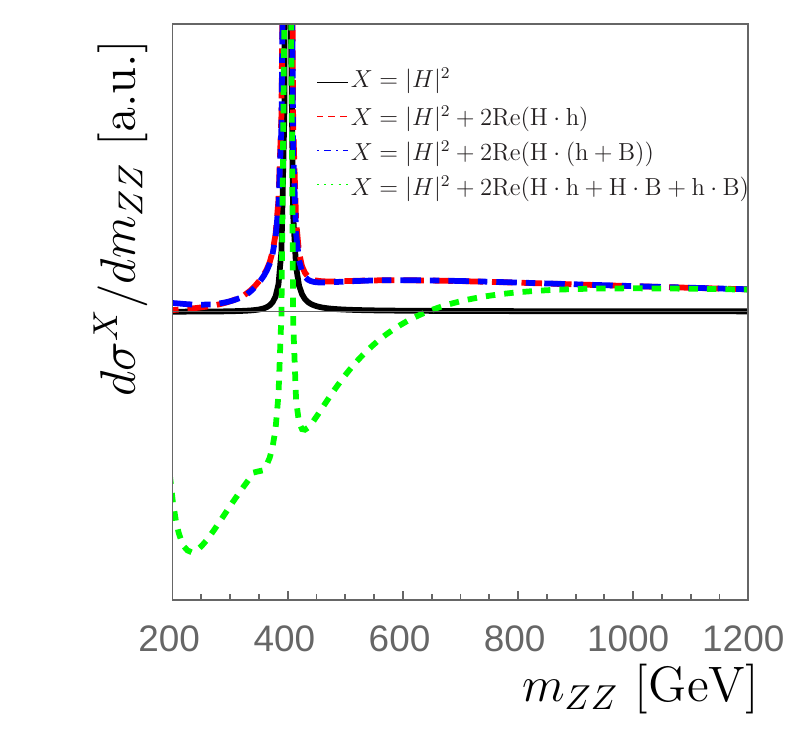}} \\[-0.7cm]
\multicolumn{2}{c}{(c)} 
\end{tabular}
\end{center}
\vspace{-0.6cm}
\caption{Partonic cross sections $d\sigma^X/d\mzz$ as a function of the invariant mass $\mzz$ in GeV
for scenario (a) S2, (b) S3 and (c) S5 
(black: $X=|H|^2$; red, dashed: $X=|H|^2+2\text{Re}(H\cdot h)$;
blue, dot-dashed: $X=|H|^2+2\text{Re}(H\cdot h)+2\text{Re}(H\cdot B)$; green, dotted: $X=|H|^2+2\text{Re}(H\cdot h)+2\text{Re}(H\cdot B)
+2\text{Re}(h\cdot B)$).}
\label{fig:S246hb} 
\end{figure}
So far we focused on the interference effects between the heavy Higgs and the background 
as well as the heavy Higgs and the light Higgs in the vicinity of the heavy Higgs resonance.
Within the region of the heavy Higgs mass peak the interference between the light Higgs
and the background can be considered constant, with a negative contribution
to $d\sigma/d\mzz$,
and was therefore not considered in the discussion of $gg\rightarrow ZZ$ so far. However,
similar to our discussion of the processes with four fermionic final states
we now add the interference of the light Higgs boson with background diagrams also to $gg\rightarrow ZZ$,
since at high invariant masses the
interplay between all three contributions, $h$ and $H$ and the background~$B$,
is of relevance and plays a role in the unitarization of the cross section.
In \fig{fig:S246hb} we plot the differential cross 
section as a function of the invariant mass of the diboson system up to high masses beyond the heavy
Higgs resonance. We exemplify the discussion for the three scenarios S2, S3 and S5. The differences
between the colored curves display the importance of the different interference terms.
Since again our study is performed for the partonic cross section and we are interested in
the relative effects of the interferences among each other, we do not display units for
$d\sigma/d\mzz$. At high invariant masses the interference between the heavy Higgs boson
and the background is negligible, in contrast to the interference of the light Higgs and the
heavy Higgs boson, which can be large and can have either sign.
Moreover, the interference of the light Higgs boson and the 
background has a sizable impact below and above $\mzz=\mH$ up to invariant masses of about $1$\,TeV.
It should be noted that the interference effects below 
$\mzz=\mH$ are not easy to distinguish from the larger backgrounds in this
region and are also reduced by our selection cuts.
In the decoupling limit $|\sba|\to 1$ the interference contributions of
the light Higgs boson are of course larger than the interference $H\cdot B$.
\fig{fig:S246hb} depicts different cases where the interference $h\cdot H$ 
is either negative, similar to the interference $h\cdot B$, or leads to a positive contribution to
the differential cross section
in a region $\mzz\in\left[450\,\text{GeV},1000\,\text{GeV}\right]$.
The latter case is realised for scenarios~S3 and S5, where 
the sum of the contributions entering with different sign gives rise to a 
``peak''-like structure.
This structure also appears in the total four particle final state,
where the gluon luminosities further suppress the cross section at high
invariant masses. As a consequence, 
all interferences need to be taken into account in order to properly model 
``peak''-like structures of this kind and thus to correctly describe the cross section
at high invariant masses.

\section{Conclusions}

\label{sec:concl}

We have investigated the production of a heavy Higgs boson of a \cp{}-conserving
Two-Higgs-Doublet-Model in gluon fusion and its subsequent decay into
a four-fermion final state. We have discussed in this context
the invariant mass and transverse
mass distributions for $gg\rightarrow e^{+}e^{-}\mu^{+}\mu^{-}$ 
and $gg\rightarrow e^{+}e^{-}\nu_l\bar\nu_l$, respectively, as well
as other kinematical observables like the separation
between outgoing same-flavor leptons
and the transverse momentum distributions of the
hardest electron/positron. The analysis has been carried out for
five different benchmark scenarios. The relative importance of the interference contributions
between the heavy Higgs boson, the light Higgs boson and the background
has been investigated for the process in the on-shell approximation,
$gg\rightarrow ZZ$. 
The employed code {\tt GoSam} makes it possible to consistently take into
account all mentioned interferences for the four-fermion final states,
which should be of interest for future heavy Higgs boson searches.

We have shown that the interference effects are
essential for a correct description of the differential cross section,
in particular at high invariant masses of the gauge boson system.
Rather than being a trivial function of the heavy Higgs mass~$\mH$
and the heavy Higgs width~$\GaH$, the mentioned interferences are in particular of relevance in
regions of the parameter space where the heavy Higgs
boson signal is diminished by small couplings. In case of a \thdm{} of type~II the enhancement
of the bottom-Yukawa coupling for large values of $\tan\beta$ can also
significantly enlarge the interference effects.
We have investigated the approximation made in the recent ATLAS
analysis in the \thdm\ to neglect the interference contributions of the 
heavy Higgs boson with the background and the light Higgs boson. We have
found that 
the relative importance of those interference contributions is 
at the level of $\mathcal{O}(10\%)$ with respect to the heavy Higgs boson 
signal cross section, and that the approximation to neglect the interference
contributions involving the heavy Higgs boson was 
indeed justified in view of the experimental sensitivity
that has been reached in the first run of the \lhc. 
We have pointed out, however, that regions of the differential cross
sections where interference effects are much larger are potentially in reach 
at high integrated luminosities. 
As an important result in this context we have found that interference
contributions can significantly enhance the sensitivity to the heavy Higgs 
boson in experimental searches.
In the vicinity of the heavy Higgs boson resonance the interference
contributions are particularly important
since they simultaneously alter the form and the position of the
heavy Higgs boson mass peak. 
We have furthermore pointed out that the interference $h\cdot H$, which can
enter with either sign, in combination with other contributions can actually
mimic a ``peak''-like structure. An accurate modelling of effects of this
kind, which requires the proper incorporation of all interference
contributions, is clearly highly relevant for future searches at high
invariant masses.

\section*{Acknowledgments}

The authors acknowledge support by
``Deutsche Forschungsgemeinschaft'' through
the SFB 676 ``Particles, Strings and the Early Universe''.
This research was supported in part by the European Commission through
the ``HiggsTools'' Initial Training Network PITN-GA-2012-316704.
NG was supported by the Swiss National Science Foundation under contract PZ00P2\_154829.
We acknowledge the DESY Bird/NAF Cluster and the Helmholtz Alliance ``Physics at the Terascale''
for providing resources for parts of our numerical results.

%bibliography:

\end{document}